\documentclass[a4paper]{aa}
\usepackage{graphicx}
\usepackage{latexsym,amsmath,amssymb}
\usepackage{natbib}
\bibpunct{(}{)}{;}{a}{}{,}

\def \xmm {XMM-Newton}
\def \src {XB\,1254$-$690}
\def \degmark{^\circ}

\def \hcm {\hbox {\ifmmode $ atom cm$^{-2}\else atom cm$^{-2}$\fi}}
\def \arcmin {\hbox{$^\prime$}}
\def \arcsec {\hbox{$^{\prime\prime}$}}
\def \deg {$^{\circ}$}
\def \chisq {$\chi ^{2}$}
\def \rchisq {$\chi_{\nu} ^{2}$}
\def\approxgt{\mathrel{\hbox{\rlap{\lower.55ex \hbox {$\sim$}}
        \kern-.3em \raise.4ex \hbox{$>$}}}}
\def\approxlt{\mathrel{\hbox{\rlap{\lower.55ex \hbox {$\sim$}}
        \kern-.3em \raise.4ex \hbox{$<$}}}}

\newcommand{\mc}{\multicolumn}

\newcommand {\oone} {\ion{O}{i}}
\newcommand {\otwo} {\ion{O}{ii}}
\newcommand {\othree} {\ion{O}{iii}}

\newcommand {\fetsix} {\ion{Fe}{xxvi}}

\newcommand {\oeight} {\ion{O}{viii}}

\newcommand {\nenine} {\ion{Ne}{ix}}
\newcommand {\neten} {\ion{Ne}{x}}

\newcommand {\mgtwelve} {\ion{Mg}{xii}}

\def\ang {$\rm\AA$}

\def\ergcmsec{\hbox{erg cm$^{-2}$ s$^{-1}$}}
\def\countsec{\hbox{count s$^{-1}$}}

\def\degmark{^\circ}

\def \nineteen {XB\,1916$-$053}
\def \mxb {MXB\,1658$-$298}

\def \seventeen {H\,1743$-$322}
\def \threethreenine {GX\,339$-$4}
\def \xte {XTE\,J1650$-$500}

\def \nhwarmabs {$N{\rm _H^{warmabs}}$}

\def \xiunit {\hbox{erg cm s$^{-1}$}}
\def \logxi {$\log(\xi)$}

\def\countsec{\hbox{counts s$^{-1}$}}

\newcommand {\egau} {$E_{\rm gau}$}

\newcommand {\ktbb} {$kT_{\rm bb}$}
\newcommand {\ktcomptt} {$kT_{\rm comptt}$}
\newcommand {\kcomptt} {$k_{\rm comptt}$}

\newcommand {\ew} {$EW$}

\def \nhabs {$N{\rm _H^{abs}}$}

\def \xiunit {\hbox{erg cm s$^{-1}$}}
\def \logxi {$\log(\xi)$}

\newcommand {\sigmav} {$\sigma_{\rm v}$}
\newcommand {\kms} {km~s$^{-1}$}

\def \kbb {$k_{\rm bb}$}
\def \kgau {$k_{\rm gau}$}

\begin{document}

\title{Variations in the dip properties of the low-mass 
X-ray binary \src\ observed with XMM-Newton and INTEGRAL}

\author{M. D{\'i}az Trigo\inst{1} \and A.N. Parmar\inst{2} \and L. Boirin\inst{3} \and C. Motch\inst{3} \and A. Talavera\inst{1} \and S. Balman\inst{4}}
\institute{
       XMM-Newton Science Operations Centre, 
  Science Operations Department,
ESAC, P.O. Box 78, E-28691 Villanueva de la Ca\~nada, Madrid, Spain    
       \and
       Astrophysics Mission Division, Research and Scientific Support
       Department of ESA, ESTEC,
       Postbus 299, NL-2200 AG Noordwijk, The Netherlands
\and
       Observatoire Astronomique de Strasbourg, 11 rue de l'Universit\'e,
       F-67000 Strasbourg, France
\and
       Middle East Technical University, In{\"o}n{\"u} Bulvari, Ankara 06531, Turkey
}

\date{Received ; Accepted:}

\authorrunning{D{\'i}az Trigo et al.}

\titlerunning{\src\ dip properties}

\abstract{We have analysed data from five \xmm\ observations of \src, 
one of them simultaneous with INTEGRAL, to investigate the mechanism
responsible for the highly variable dips durations and depths seen from
this low-mass X-ray binary.  Deep dips were present during two
observations, shallow dips during one and no dips were detected
during the remaining two observations. At high (1--4~s) time
resolution ``shallow dips'' are seen to include a few, very
rapid, deep dips whilst the ``deep'' dips 
consist of many similar very rapid, deep, fluctuations.  
The folded V-band Optical
Monitor light curves obtained when the source was undergoing deep,
shallow and no detectable dipping exhibit sinusoid-like variations
with different amplitudes and phases. We fit EPIC
spectra obtained from "persistent" or dip-free intervals with a model
consisting of disc-blackbody and thermal comptonisation components
together with Gaussian emission features at 1 and 6.6~keV modified by
absorption due to cold and photo-ionised material. None of the
spectral parameters appears to be strongly correlated with the dip
depth except for the temperature of the disc blackbody which is
coolest (kT$\sim$1.8~keV) when deep dips are present and warmest
(kT$\sim$2.1~keV) when no dips are detectable. We propose that the
changes in both disc temperature and optical modulation could be
explained by the presence of a tilted accretion disc in the
system. 
We provide a revised estimate of the orbital period of
$0.16388875 \pm 0.00000017$~day. } \maketitle

\section{Introduction}
\label{sect:intro}
Around 10 galactic low-mass X-ray binaries (LMXBs) exhibit
periodic dips in their X-ray intensity. The dips 
are believed to be caused by
periodic obscuration of the central X-ray source by structure
located in the outer regions of a disc
\citep{1916:White82apjl}. The depth, duration and spectral
evolution of the dips varies from source to source and often from cycle
to cycle. The 1--10~keV spectra of most of the dip sources become
harder during dipping. However, these changes are inconsistent
with a simple increase in photo-electric absorption by cool
material, as an excess of low-energy photons is usually present.
Narrow absorption
features from highly ionised Fe and other metals have been
observed from a 
number of dipping LMXBs and microquasars \citep[e.g.,][]{1655:ueda98apj,
1658:sidoli01aa, 1915:lee02apj}. The important
role that photo-ionised plasmas play in LMXBs was recognised by
\citet{1323:boirin05aa} and \citet{ionabs:diaz06aa} who were able
to model the changes in {\it both} the narrow
X-ray absorption features and the continuum during the dips from all the bright
dipping LMXB observed by XMM-Newton by an increase in the column
density and a decrease in the amount of ionisation of a
photo-ionised absorbing plasma. Since dipping sources are
normal LMXBs viewed from close to the orbital plane, this implies
that photo-ionised plasmas are common features of LMXBs. Outside
of the dips, the properties of the absorbers do not vary strongly
with orbital phase suggesting that the ionised plasma 
has a cylindrical geometry with a maximum column density close to
the plane of the accretion disc.

Dipping activity from the LMXB
\src\ was discovered during EXOSAT observations in 1984 when the source
exhibited irregular reductions in X-ray intensity that
repeated every $3.88 \pm 0.15$~hr \citep[][]{1254:courvoisier86apj}.
In a 1984 August observation, five deep dips were observed with
a mean duration of $\sim$0.8~hr 
and reduction in 1--10~keV intensity of $\sim$95\%.  V-band observations
of the 19th magnitude companion identified by
\citet{1254:griffiths78nat} revealed an optical modulation with
minima occurring $\sim$0.2 cycles after the X-ray dips
\citep[][]{1254:motch87apj}.  The optical modulation has a 
period of $3.9334 \pm 0.0002$~hr, consistent with the mean X-ray dip recurrence interval, 
and can be modelled as resulting
primarily from viewing different aspects of the X-ray
heated atmosphere of the companion star with a small contribution
from the bulge where the accretion stream impacts the
outer disc 
\citep[][]{1254:motch87apj}. This indicates that the accretion disc
does not entirely shadow the companion.
The presence of dips and the lack of X-ray eclipses provides a constraint
on the inclination angle, $i$, of the source of between 65$\degmark$ and 73$\degmark$
\citep[][]{1254:courvoisier86apj, 1254:motch87apj}.

Deep dips were present during a
{\it Ginga} observation of \src\ in 1990 \citep{1254:uno97pasj}, but were
not detected during Rossi X-ray Timing Explorer (RXTE) observations
in 1997 \citep{1254:smalewachter99apj} and $BeppoSAX$ observations
in 1998 \citep{1254:iaria01apj}. 
Optical observations in 1997 revealed that
the mean V magnitude was unchanged, but that the amplitude of
the optical variability had declined by $\Delta$V $\sim$0.1 mag
\citep{1254:smalewachter99apj}.
This may be explained if the vertical structure at the outer regions of the disc responsible
for the dips decreased in angular size from 17$\degmark$--25$\degmark$
to $<$10$\degmark$  \citep{1254:smalewachter99apj}.  The dips had re-appeared
during XMM-Newton and RXTE observations in 2001 
January and May, but were again not present
during observations in 2001 December \citep[][]{1254:smale02apj}, 2002
February \citep{1254:boirin03aa} and 2003 October \citep{1254:iaria07aa}. 
Deep dipping had returned in an RXTE observation in
2004 May \citep{1254:barnes07mn}.
While all the well studied LMXB dippers show variability from cycle to cycle, 
the
large variations in dip depth observed from \src\ are unusual.
 
Here, we report the results of three XMM-Newton observations of \src\
in 2006 and 2007 as well as a  re-analysis of the two earlier 
XMM-Newton observations reported in \citet{1254:boirin03aa}
and \citet{ionabs:diaz06aa}. The new observations include results
from the Optical Monitor which was not operated in the early observations.
The second of the new
observations was simultaneous with an INTEGRAL observation which is used
to extend the analysis to higher energies.
The goal of this work is to investigate and explain the mechanism
presumed responsible for the large changes in dip depth.  

\section{Observations}
\label{sec:observations}

\subsection{XMM-Newton observations}
The XMM-Newton Observatory \citep{xmm:jansen01aa} includes three
1500~cm$^2$ X-ray telescopes each with an EPIC 
(0.1--15~keV) at the focus. Two of the EPIC imaging
spectrometers use MOS CCDs \citep{xmm:turner01aa} and one uses pn CCDs
\citep{xmm:struder01aa}. The RGSs \citep[0.35--2.5~keV,][]{xmm:denherder01aa} 
are located behind two of the
telescopes. In addition, there is a co-aligned 30~cm diameter 
Optical/UV Monitor telescope  \citep[OM,][]{xmm:mason01aa}, 
providing simultaneously coverage with the X-ray instruments.
Data products were reduced using the Science Analysis
Software (SAS) version 7.1. Since the EPIC pn is more
sensitive to the presence of Fe lines than the MOS CCDs, with an
effective area a factor $\sim$5 higher at 7~keV, we concentrate here 
on the analysis of pn data from EPIC.
RGS data from both gratings in first and second order were also
used as were OM data from the three available XMM-Newton observations.

\begin{table*}
\begin{center}
\caption[]{XMM-Newton observations of \src. Results from the first two
observations are reported in \citet{1254:boirin03aa} and
\citet{ionabs:diaz06aa}. $T$ is the total EPIC pn exposure time, $C$ 
is the pn 0.6--10~keV persistent emission background subtracted
count rate and $HR$ the Hardness Ratio (counts in the 2--10~keV energy
range divided by those between 0.6--2~keV) for the dip free
intervals. Obs 4 was simultaneous with INTEGRAL. In all cases the
EPIC pn thin filter was used. "Deep" is a reduction of $\sim$10--90\%
in the 0.6--10~keV X-ray flux within 1~hr around the dip centre
compared to the flux within 1~hr measured 2~hrs away from the dip
centre, "Shallow" is a reduction of $\sim$5--10\% and "Undetected"
implies a reduction of $\approxlt$5\%.}
\begin{tabular}{cclcccccc}
\hline \noalign {\smallskip}
Obs  & Observation & \mc{3}{c}{Observation Times (UTC)} & Dip & $T$  & $C$ & $HR$   \\
Num & ID   & \mc{2}{c}{Start}  & End & Depth   & (ks) & (s$^{-1}$) & \\
        &        & (year~month~day)& (hr:mn) & (hr:mn) \\
\hline \noalign {\smallskip}
1 & 0060740101 & 2001 January   22 & 15:49 & 20:03 & Deep       & 15.0 & 170.8 $\pm$ 0.2 & 0.691 $\pm$ 0.001 \\
2 & 0060740901 & 2002 February  7  & 17:32 & 25:08 & Undetected & 26.8 & 178.2 $\pm$ 0.2 & 0.770 $\pm$ 0.001 \\
3 & 0405510301 & 2006 September 12 & 16:15 & 08:01 & Undetected & 56.0 & 192.5 $\pm$ 0.2 & 0.7665 $\pm$ 0.0006 \\
4 & 0405510401 & 2007 January   14 & 01:13 & 18:20 & Deep       & 60.8 & 178.3 $\pm$ 0.2 & 0.7044 $\pm$ 0.0005 \\
5 & 0405510501 & 2007 March     9  & 04:03 & 20:44 & Shallow    & 59.2 & 183.0 $\pm$ 0.2 & 0.7267 $\pm$ 0.0007 \\
\noalign {\smallskip} \hline \label{tab:obslog}
\end{tabular}
\end{center}
\end{table*}

\begin{figure*}[!ht]
\includegraphics[angle=0.0,height=0.2\textheight]{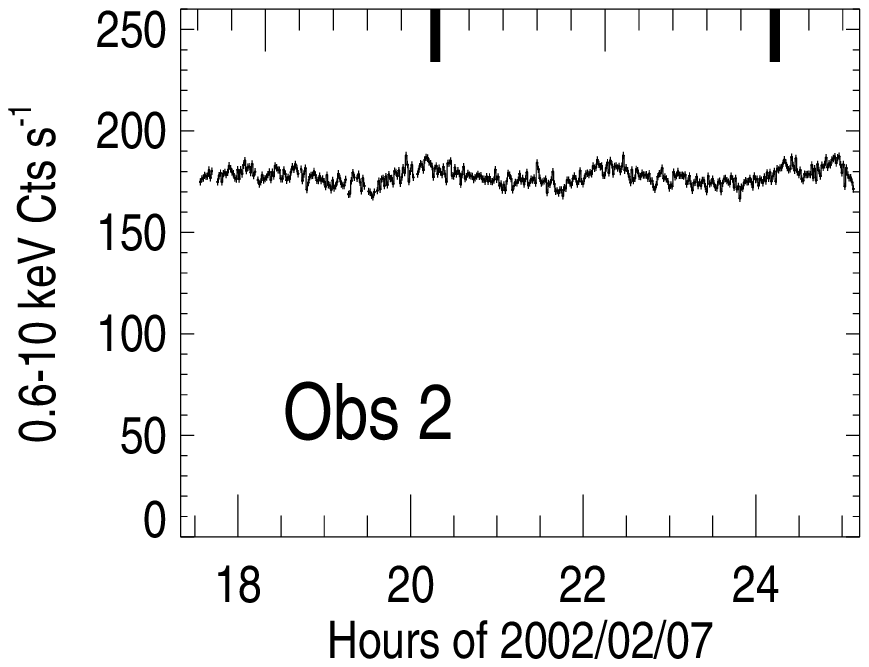}
\includegraphics[angle=0.0,height=0.2\textheight]{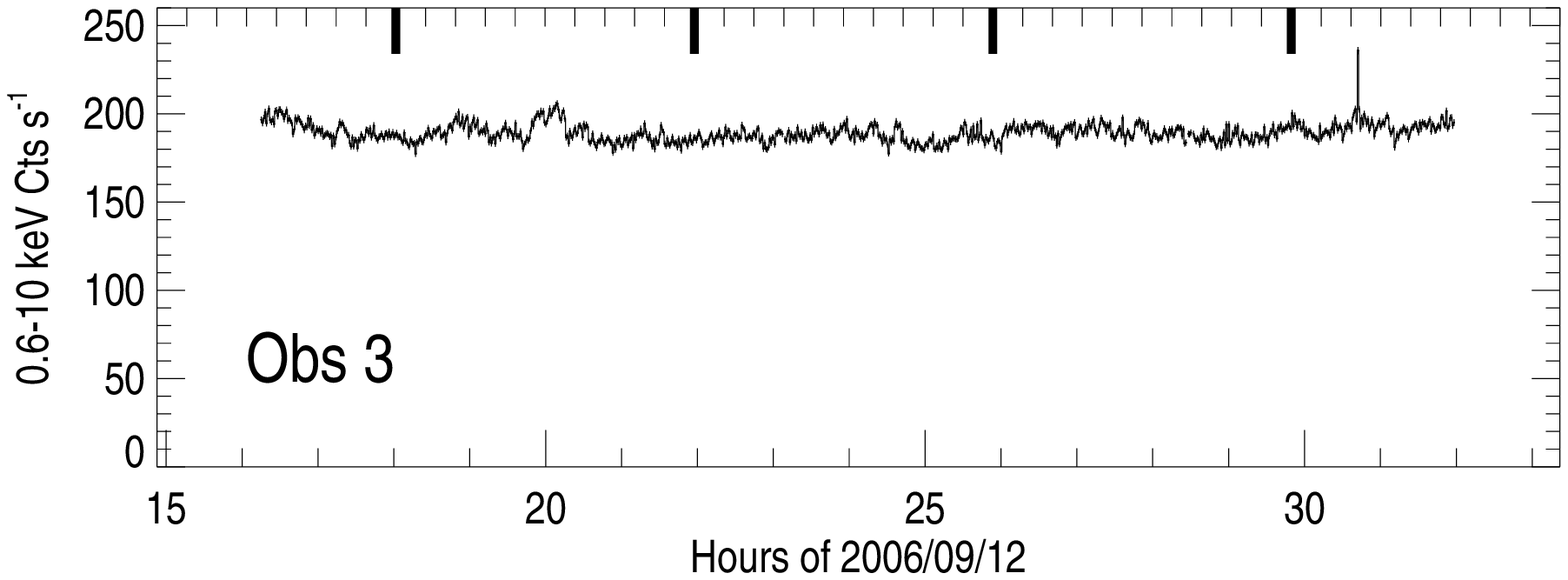}\\
\includegraphics[angle=0.0,height=0.2\textheight]{fig1-3.ps}
\includegraphics[angle=0.0,height=0.2\textheight]{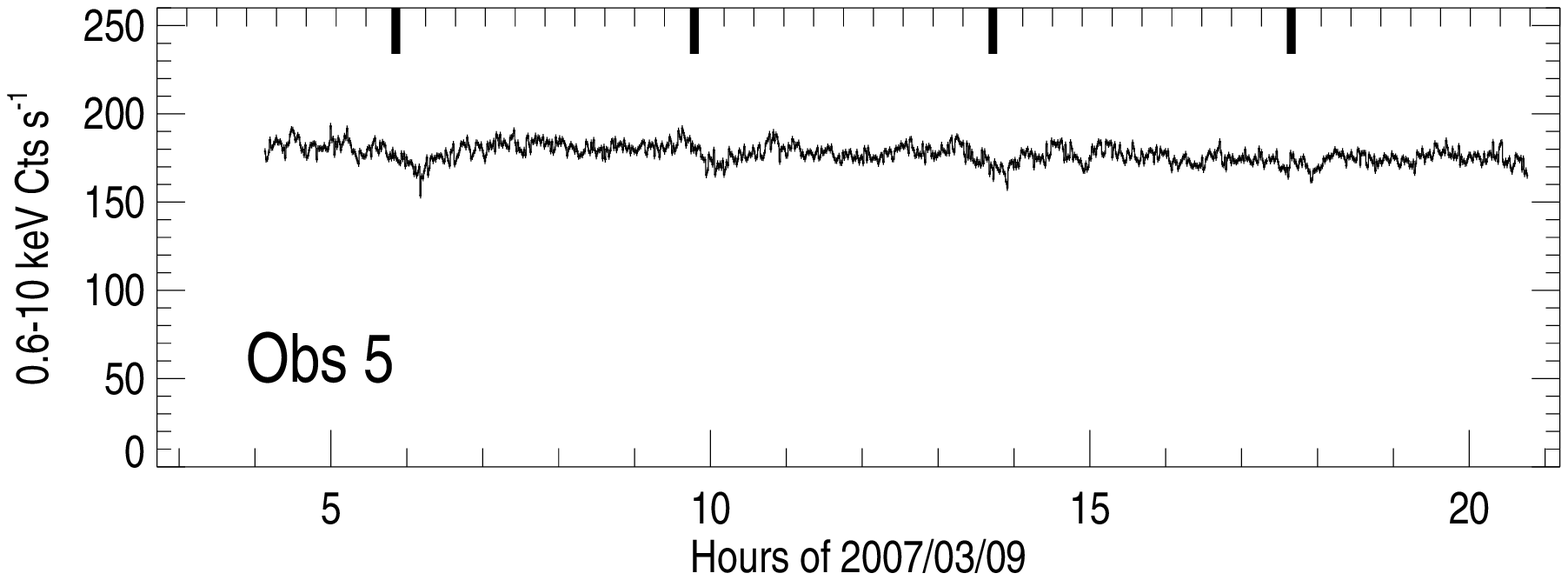}\\
\includegraphics[angle=0.0,height=0.2\textheight]{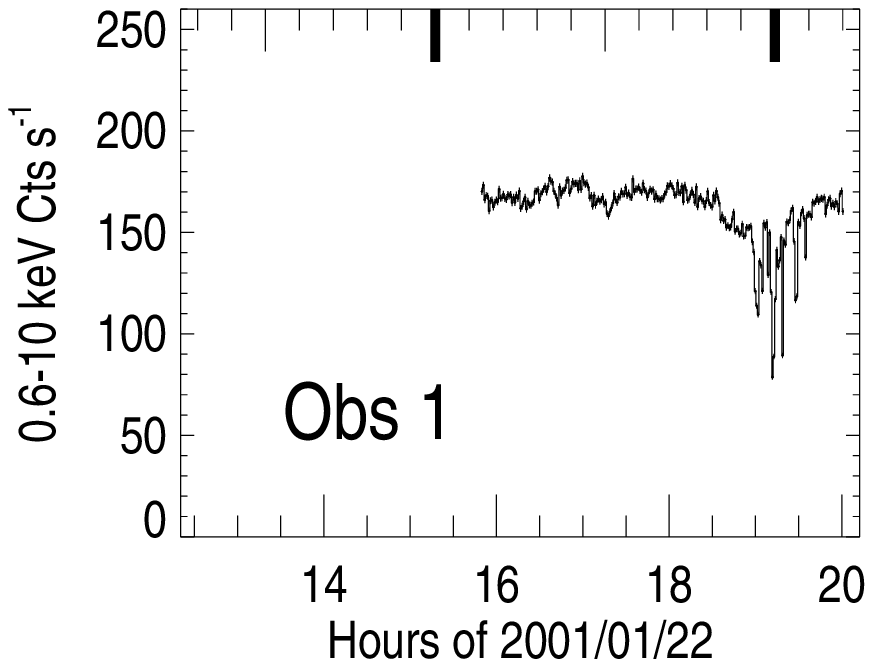}
\includegraphics[angle=0.0,height=0.2\textheight]{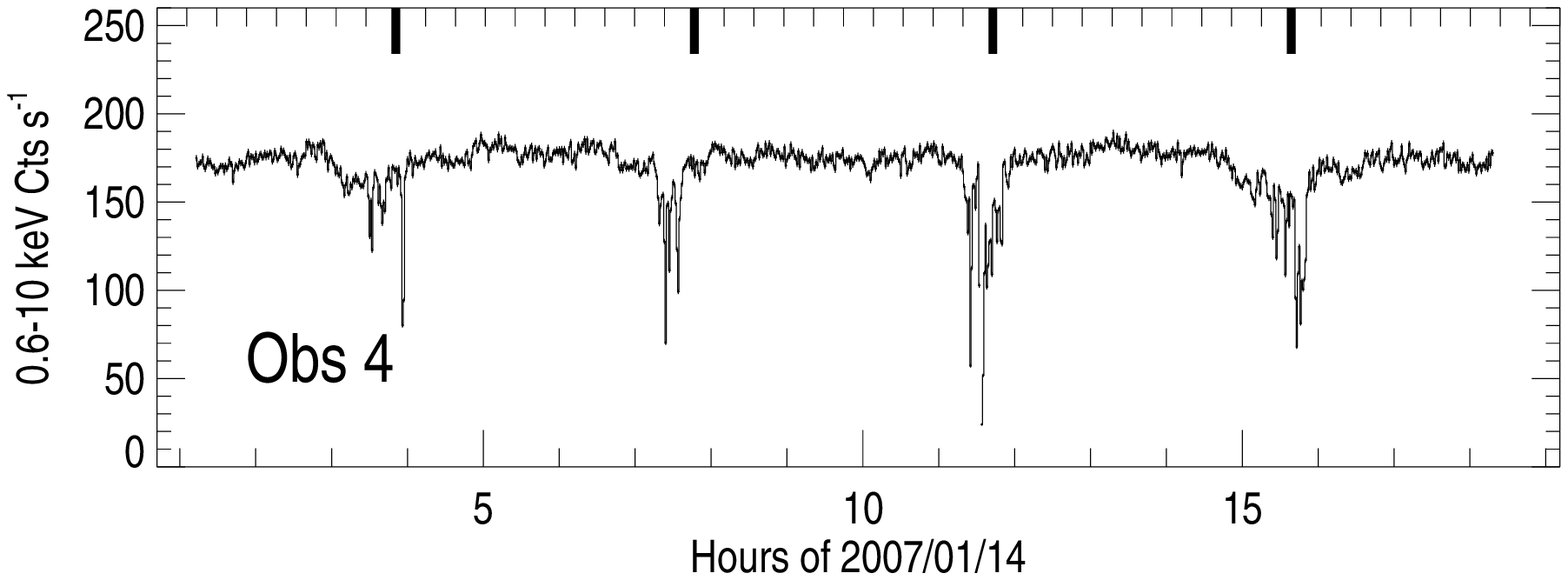}\\
\vspace{-0.5cm}
\caption{0.6--10 keV EPIC pn background subtracted light curves for
each \src\ observation with a binning of 120~s. The panels have been
arranged so that the observations not showing dips (Obs~2 and~3) are
shown at the top, the observation showing shallow dips (Obs~5) in the middle
and the observations showing deep dips (Obs~1 and~4) at the bottom.
The axes scales are the same for all observations. The thick vertical
tick marks indicate the expected dip centres using the period of
optical variations given in \citet{1254:motch87apj} and using the deep
dips in Obs~4 to define the dip phase (see text). A burst is visible
in Obs~3.}
\label{fig:lightcurves}
\end{figure*}

Table~\ref{tab:obslog} is a summary of the XMM-Newton observations. We
used the EPIC pn in Timing Mode. In this mode only one CCD chip is
operated and the data are collapsed into a one-dimensional row
(4\farcm4) and read out at high speed, the second dimension being
replaced by timing information. This allows a time resolution of
30~$\mu$s, and photon pile-up occurs only for count rates
$>$800~\countsec, well in excess of the observed count rates (see
Fig.~\ref{fig:lightcurves}). Only single and double Timing Mode
events (patterns 0 to 4) were selected and source events were
extracted from a 53\arcsec\ wide column centred on the source
position. Background events were obtained from a column of the same
width, but positioned away from the source. Ancillary response files
were generated using the SAS task {\tt arfgen}. Response matrices were
generated using the SAS task {\tt rmfgen}.  The SAS task {\tt rgsproc}
was used to produce calibrated RGS event lists, spectra, and response
matrices. We chose the option {\it keepcool}=no to discard single
columns that give signals a few percent below the values expected from
their immediate neighbours. Such columns are likely to be important
when studying weak absorption features in spectra with high
statistics. The OM was operated in Image+Fast Mode in Obs~3--5. The V
filter, very close to the standard Johnson V band, was used. In this
mode the instrument produces images of the entire 17\arcmin\ x
17\arcmin\ FOV with a time resolution between 800 and 5000~s and event
lists with a time resolution of 0.5~s from a selected 11\arcsec\ x
11\arcsec\ region. The SAS task {\tt omfchain} was used to extract
light curves of \src\ from the high time resolution Fast Mode data. We
used a time sampling of 100~s in the curve extraction to improve the
signal-to-noise ratio. We corrected the OM Fast Window light curves
and EPIC pn event times to the solar system barycentre using the SAS
task {\tt barycen}.

\begin{figure*}[!ht]
\includegraphics[angle=90,width=0.49\textwidth]{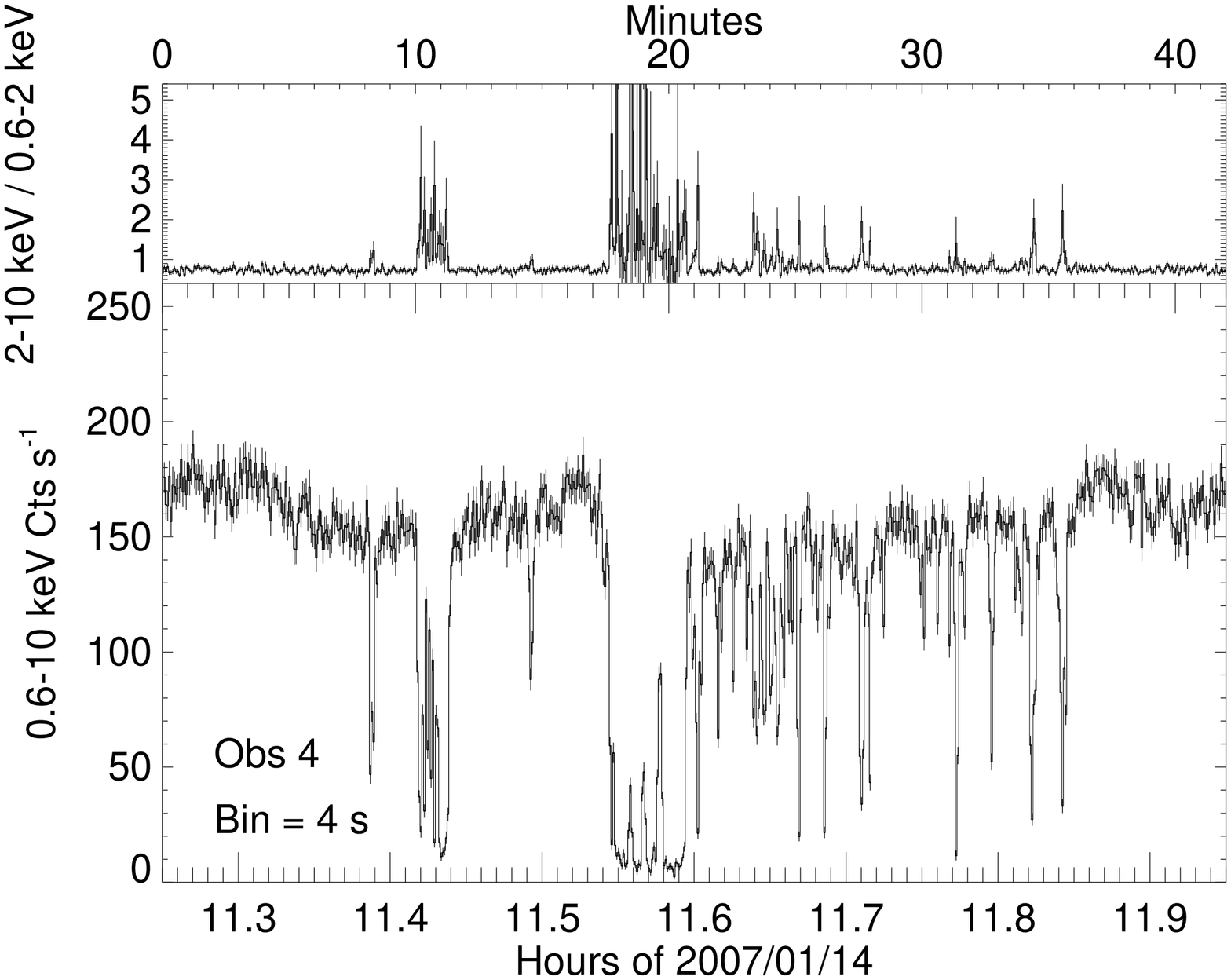}
\includegraphics[angle=90,width=0.49\textwidth]{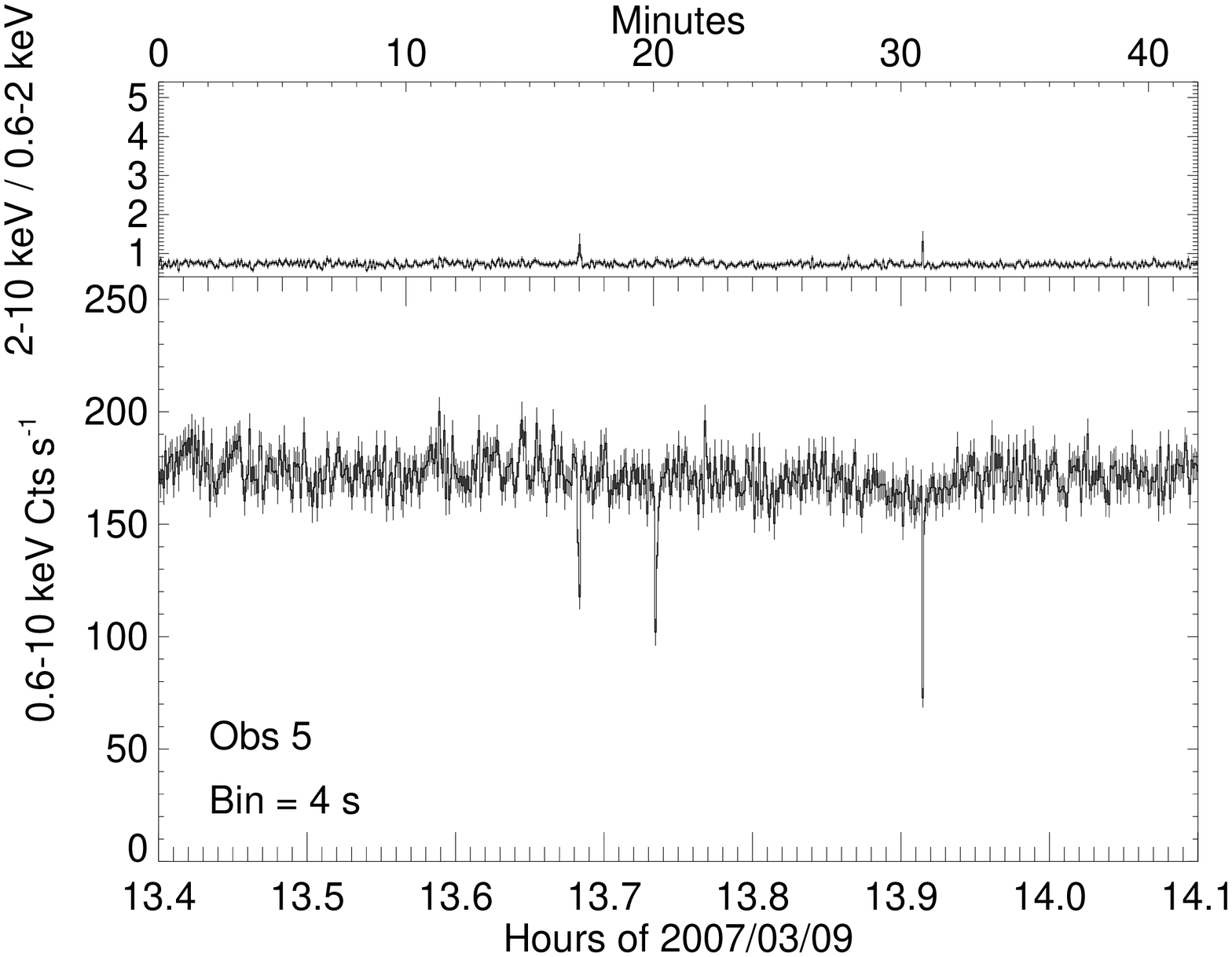}
\includegraphics[angle=90,width=0.49\textwidth]{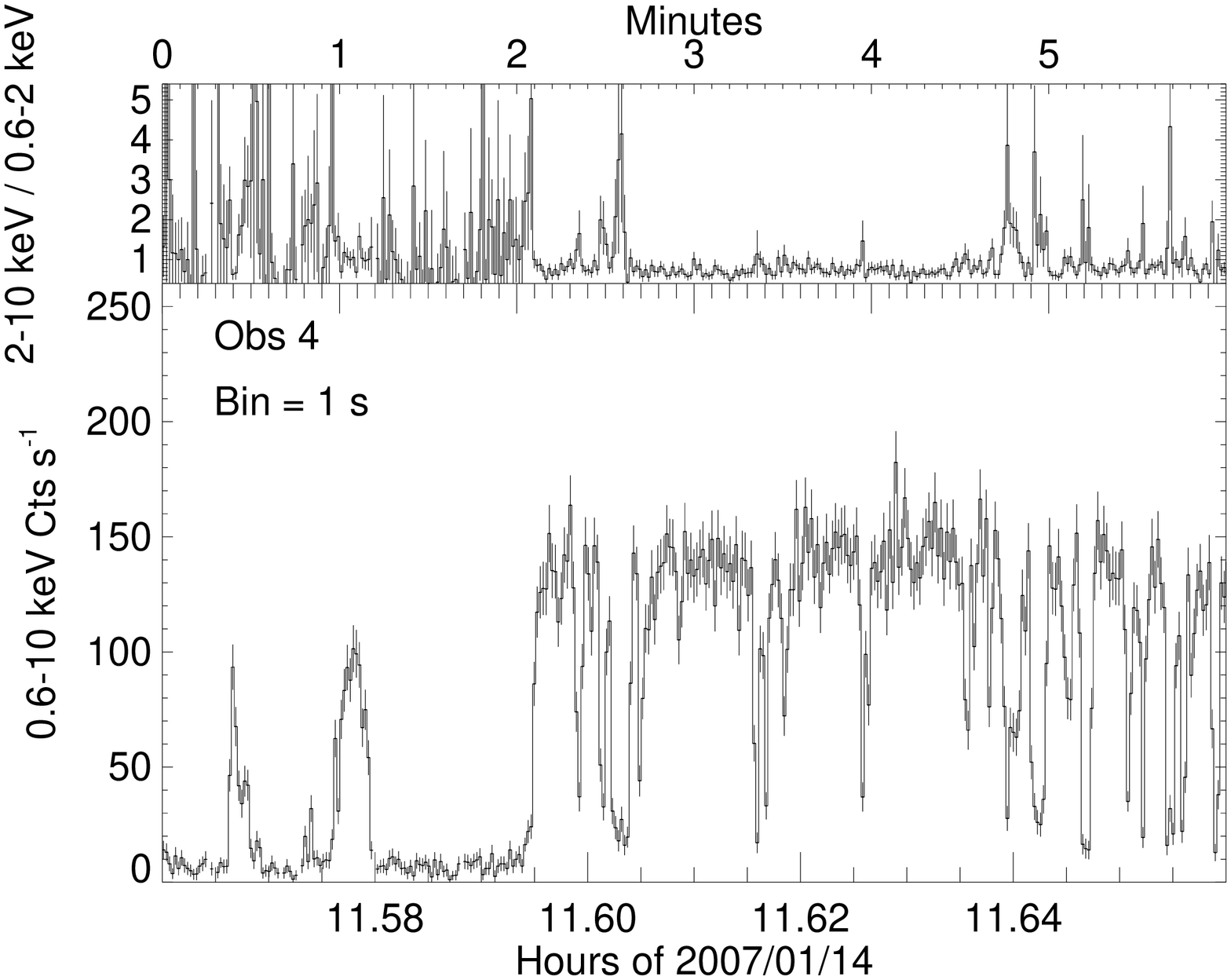}
\includegraphics[angle=90,width=0.49\textwidth]{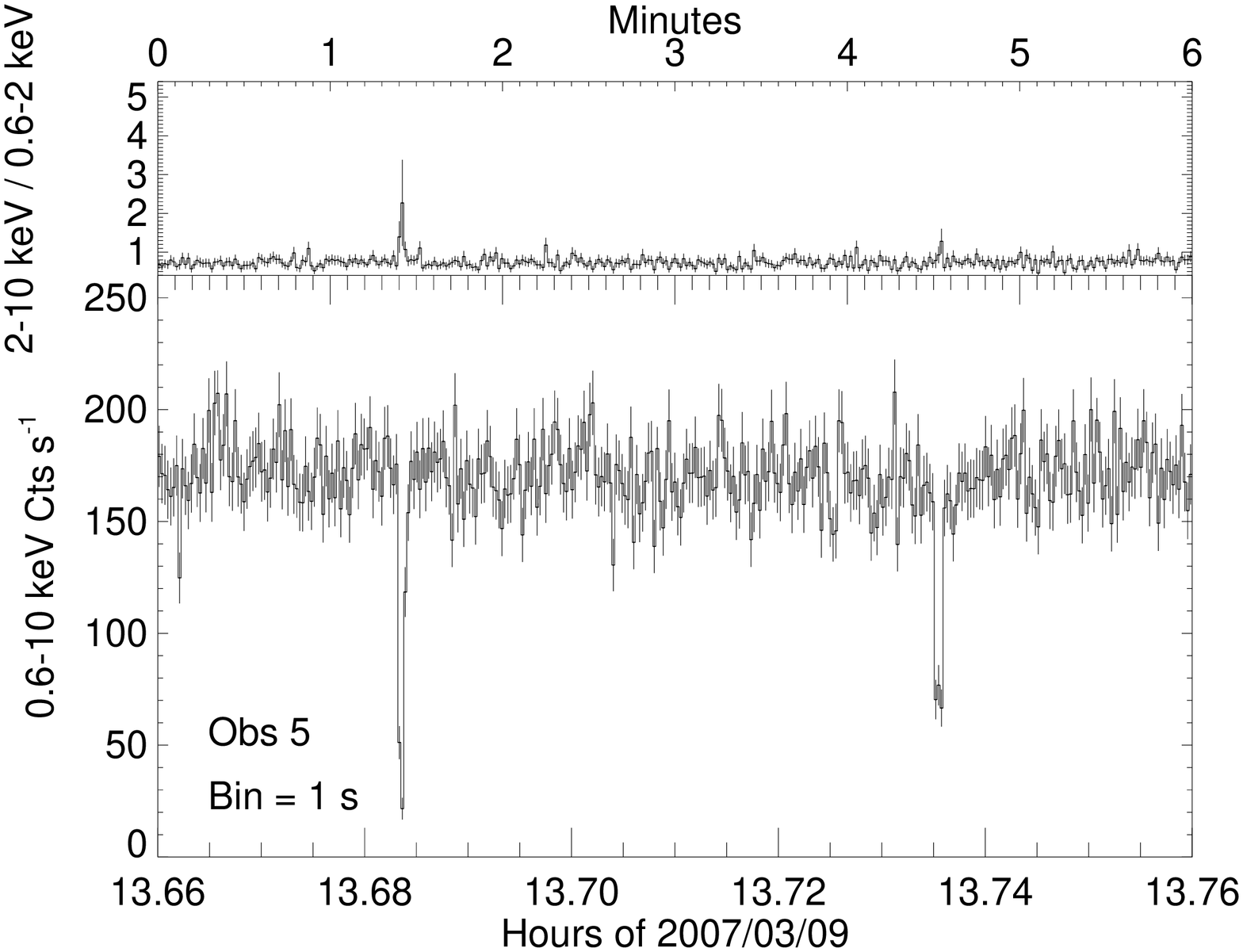}
\caption{EPIC pn 0.6--10 keV light curves and hardness ratios for
\src, with a binning of 4~s (upper panels) and 1~s (lower panels),
showing many, rapid, and deep intensity fluctuations during a deep dip
(left). The "shoulder" (see definition in Sect.~\ref{sec:x-lc}) is visible between 11.3 and 11.4~hrs in the upper left panel. In contrast, the shallow dip (right), 
when plotted at high
time resolution, is seen to consist of a smaller number of similar
rapid deep intensity fluctuations. The hardness ratio is counts in the
2--10~keV energy range divided by those between 0.6--2~keV.}
\label{fig:detail-lightcurves}
\end{figure*}

\subsection{INTEGRAL observation}
The INTEGRAL \citep{Winkler03AA} payload consists of two main
gamma-ray instruments, one of which is optimized for 15~keV to 10~MeV
high-resolution imaging \citep[IBIS;][]{Ubertini03AA}. 
IBIS provides an angular resolution of
$12\arcmin$ full-width at half-maximum (FWHM) and an energy
resolution, $E/\Delta E$, of $\sim$12~FWHM at 100~keV. 
The extremely broad energy range of IBIS is
covered by two separate detector arrays, ISGRI
\citep[15--1000~keV;][]{lebrun03AA} and PICsIT
\citep[0.175--10~MeV;][]{labanti03AA}. The payload is complemented by an
X-ray monitor \citep[JEM-X; 3--35~keV;][]{Lund03AA}. 
JEM-X has a fully coded FOV of
$4\fdg8$ diameter and an angular resolution of $3\arcmin$ FWHM. 

An INTEGRAL observation of the region of sky
containing \src\ was performed between 2007 January 14 02:02 and
2007 January 15 05:47~UTC, for a total time of 74~ks, 
coinciding with \xmm\ Obs~4.
The standard 5 x 5 dither patterns of pointings centred on the
target position were performed.
Data from the observation were processed using the Off-line
Scientific Analysis (OSA) version 7.0. In this paper we
use data from ISGRI and JEM-X, both of which use
coded masks to provide imaging information. This means that photons from
a source within the FOV are distributed over the
detector area in a pattern determined by the position of the
source in the FOV. Source positions and intensities are determined
by matching the observed distribution of counts with those
produced by the mask modulation. 

\subsection{X-ray light curves}
\label{sec:x-lc}

Figure~\ref{fig:lightcurves} shows 0.6--10 keV EPIC pn light curves of
all five XMM-Newton observations of \src\ with a binning of 120~s. The
plots have been grouped so that the upper panels show the two
observations where dipping was not detected (Obs~2 and~3), the middle
panels the observation where shallow dipping was detected (Obs~5), 
and the lower panels the two observations with deep dipping
(Obs~1 and~4). "Deep" is defined as a reduction of the 0.6--10~keV flux of
$\sim$10--90\% when integrated over 1~hr around the dip centre
compared to the flux within 1~hr measured 2 hours away from the dip
centre, "Shallow" is a similar reduction of $\sim$5--10\% and
"Undetected" implies a reduction of $\approxlt$5\%. The thick
vertical tick marks indicate the expected dips times derived using the
reference epoch of JD 2,454,114.65502 and a folding period of
$0.163890 \pm 0.000009$~day \citep{1254:motch87apj}. The reference
epoch corresponds to the centre of the first X-ray dip in Obs~4 and
was determined by folding the 0.6--10~keV EPIC pn light curve of Obs~4
with the source period and fitting the resultant deep dip with a
negative Gaussian function.

\begin{figure}[!ht]
\includegraphics[angle=0,width=0.45\textwidth]{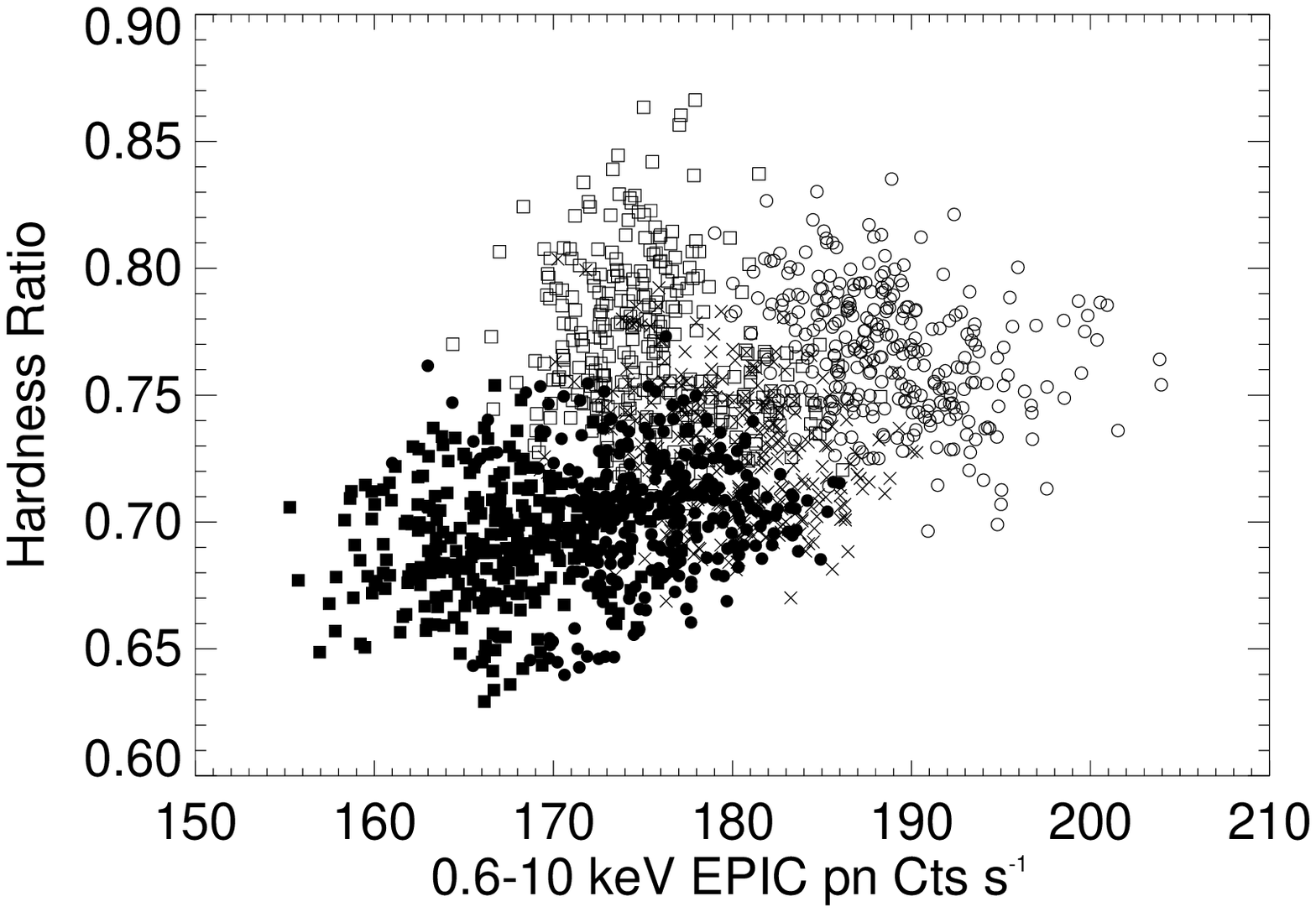}
\caption{Hardness ratio (2--10~keV/0.6--2~keV counts) versus
0.6--10~keV EPIC pn count rate during dip-free intervals. Data from
Obs~1 to 5 are shown as filled squares, open squares, open circles,
filled circles and crosses, respectively. An increase of the hardness
ratio from the observations showing deep dipping (filled symbols) to
the observations where dipping is absent (open symbols) is visible.}
\label{fig:hr}
\end{figure}

The dip shape and depth vary from dip to dip during Obs~1 and 4.
However, the dips seem to follow a regular pattern. An EPIC pn
0.6--10~keV light curve and hardness ratio (counts in the 2--10~keV
energy range divided by those between 0.6--2~keV) plot of a deep dip
with a time resolution of 4~s are shown in the upper left panels of
Fig.~\ref{fig:detail-lightcurves}. The lower left panels show part of
the same dip with a time resolution of 1~s. Examination of the upper
panels shows that the intensity first decreases slightly ("shoulder")
while the hardness ratio remains more or less unchanged, and then at some
point both the intensity and the hardness ratio become highly variable
on a wide range of timescales. The deepest segments of the dips are
associated with the strongest hardening. Light curves and hardness
ratio plots from the shallow dips seen in Obs~5 are shown in the right
panels. When plotted at the time resolution of
Fig.~\ref{fig:lightcurves} (120~s) the shallow dips appear similar to
the "shoulders" of the deep dips in the rebinned light curve. However,
at the higher 1 or 4~s time resolutions of
Fig.~\ref{fig:detail-lightcurves} it is evident that the shallow dip
also includes few very short duration deep dips with depths
and hardness ratios similar to the deep dips seen during
Obs~1 and 4.
We examined the hardness ratio as a
function of 0.6--10~keV count rate for the dip-free intervals of all
the observations (Fig.~\ref{fig:hr}).  The hardness ratio increases
from the observations showing deep dipping (Obs~1 and 4) to the one
showing shallow dipping (Obs~5) and is largest for the observations
where dipping behaviour is absent (Obs~2 and 3). The average count
rate increased from the 2000--2001 observations (Obs~1 and~2) to the 2006--2007
ones (Obs~3 to~5). A higher count rate during observations where dipping is absent
compared to observations where dipping is present is observed in both
epochs. 

\begin{figure}[!ht]
\includegraphics[angle=90,width=0.45\textwidth]{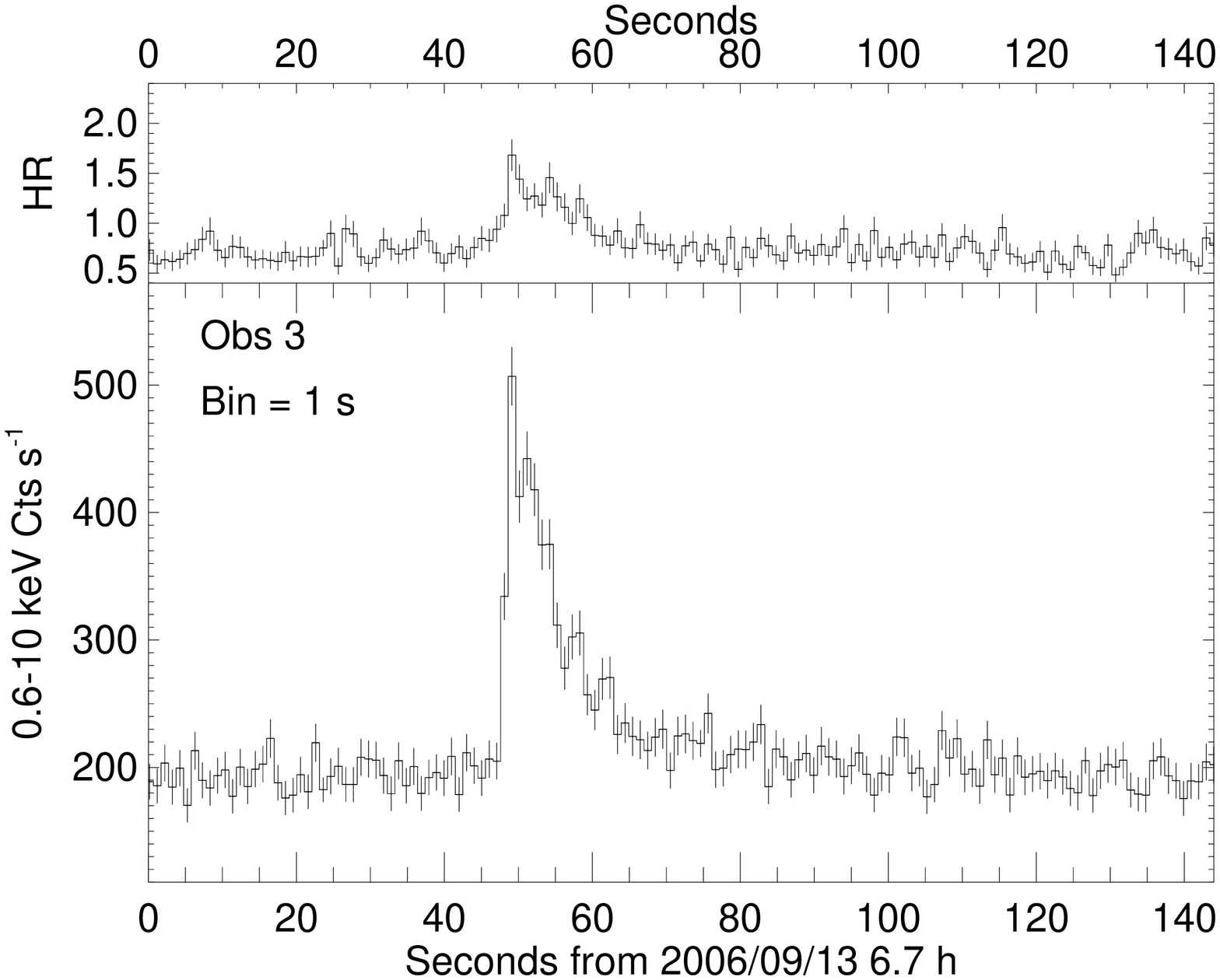}
\caption{The 0.6--10 keV EPIC pn light curve and hardness ratio 
(counts in the 2--10~keV energy range divided by those between 0.6--2~keV)
for \src\ Obs~3 during the burst with a binning of 1~s.}
\label{fig:burst-lightcurve}
\end{figure}

We note that
an X-ray burst is observed in 2006 September 13 at 6.71~hr during
Obs~3 (see Fig.~\ref{fig:burst-lightcurve}).

\subsection{Folded lightcurves}
\label{sec:op-lc}

We extracted light curves from the V filter OM Fast Window exposures
performed in Obs~3--5, with a time bin of 100~s. The OM was not operated 
during Obs~1 and~2. The
background region was carefully defined to avoid contamination by
field stars located close to the optical counterpart of \src, which
were present in the OM Fast Window region.  During some exposures in
Obs~3 the optical counterpart was located close to the edge, or even
outside, of the Fast Window region due to spacecraft pointing drift. Based on
the known OM point spread function, we rejected exposures where
$>$50\% of the source flux was outside the Fast Window region as it
was difficult to reliably reconstruct the total count rates.  The
light curves show a sinusoid-like modulation consistent with the
period of the binary. However, at a similar time resolution as the
X--ray light curve in Fig.~\ref{fig:lightcurves}, the uncertainties on
the OM lightcurves are too large to reveal details of the modulation.
Thus, we resampled the light curves at 1416~s (one tenth of the binary
period) and folded the light curves using the ephemeris presented
earlier and examined each observation separately\footnote[1]{We note that SAS 
calculates the errors for the OM light curves based on a Bernouilli
distribution rather than a Poisson distribution \citep{xmm:kuin08mnras}. 
Therefore, the Gaussian error propagation used for the folding of the
curves and the \chisq\ statistic may be not appropriate.}. 

\begin{figure*}[!ht]
\centerline{\includegraphics[width=1.\textwidth]{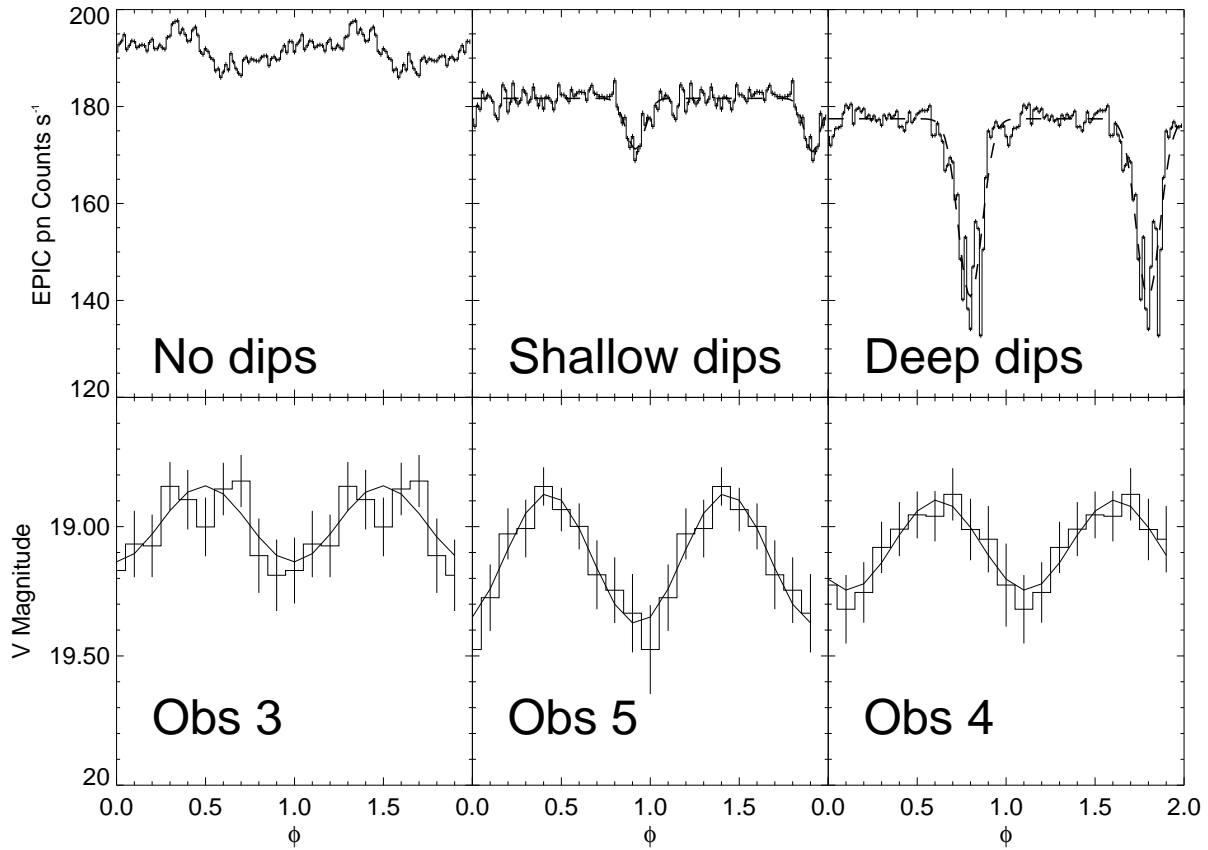}}
\caption{0.6--10 keV EPIC pn (upper panels) and V filter OM (lower
panels) \src\ light curves folded at the refined orbital period determined in Sect.~\ref{sec:lc-superperiodicity} when no dips, shallow dips and deep dips
were present. The reference epoch for phase is the same for all the
plots and $\phi$ = 0.0 corresponds to optical minimum. The reference phase
for the optical minimum has been calculated taking into account the dip 
reference phase calculated in Sect.~\ref{sec:x-lc} and assuming that the 
dip centre occurs at $\phi$ = 0.8. The binning is
1416~s and 200~s for the OM and EPIC pn light curves,
respectively. The solid lines in the lower panels show the fit of
Table~\ref{tab:omfit}.}
\label{fig:omlightcurves}
\end{figure*}

The folded light curves reveal differences in shape and magnitude
between the three observations. Figure~\ref{fig:omlightcurves} shows
V filter OM (lower panels) and simultaneous 0.6--10~keV EPIC pn (upper
panels) light curves of \src\ folded on the refined orbital period determined in Sect.~\ref{sec:lc-superperiodicity}, where $\phi$ = 0 corresponds to optical
minimum. We used the same ephemeris to fold all 3 observations. The
Obs~3 OM light curve shows evidence for a secondary minimum at
$\phi \sim$ 0.5 which may also be present in Obs~4 as an apparent 
flattening of the curve around this phase, but is clearly absent in Obs~5. The
average EPIC pn count rate decreased from Obs~3 to Obs~5. Finally, the
phases of both the optical and X-ray modulations changed between
the observations. In X-rays, we observe that the
shallow dips in Obs~5 appear $\sim$0.1 later in phase compared to the
deep dips in Obs~4. This difference is unlikely to be due to
uncertainties in the ephemeris and the period propagation since the
accumulated uncertainty from Obs~4 to Obs~5 is estimated to be $\phi
<$ 0.0005 (see Sect.~\ref{sec:lc-superperiodicity}).
Similarly, while the optical minimum falls near $\phi = 0$ for both
Obs~3 and 5, it occurs later at $\phi = 0.1$ for Obs~4, when deep dips
are present.

\begin{figure}[!ht]
\centerline{\includegraphics[angle=0,width=0.45\textwidth]{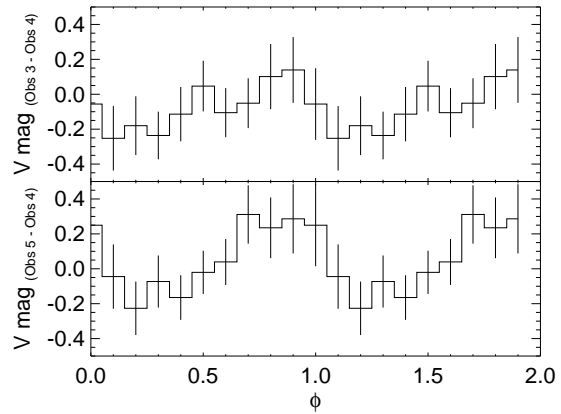}}
\caption{Differences between the OM light curves shown in
Fig.~\ref{fig:omlightcurves} (lower panels). The upper panel shows the
difference between the light curves of Obs~3 and~4 and
the lower panel between Obs~5 and~4. An excess of optical emission is seen at 
$\phi$~$\sim$~0.8 in Obs~4, when deep dipping was taking place.}
\label{fig:omlightcurves-subtracted}
\end{figure}

We modelled the OM folded light curves with a sinusoid 
with the period fixed at the orbital value. This model fits 
well the folded light curves of Obs~4 and 5. The fit quality
for Obs~3 is slightly worse than the others
due to the possible presence of a secondary
minimum at $\phi \sim 0.5$ which is not modelled.
The results of the fits are shown in the lower panels
of Fig.~\ref{fig:omlightcurves} and given in
Table~\ref{tab:omfit}. 
Finally, we subtracted the OM light curve of Obs~4 from Obs~3 and~5 
(see Fig.~\ref{fig:omlightcurves-subtracted}). An excess is observed
at $\phi \sim 0.8$ in both subtracted light curves, indicating additional optical
emission at this phase in Obs~4, when deep dipping was present, compared
to Obs~3 and~5. 

\begin{table}
\begin{center}
\caption[]{Best-fits to the folded OM light curves for Obs~3--5 using
  the {\tt $M$+$A$*sin(2$\pi(c*\phi+0.25-\phi_0$))} model. $M$, $A$, $\phi$,
  and $\phi_0$ are the average optical magnitude, amplitude, phase and phase at the optical minimum, respectively. The coefficient $c$ has been fixed to 1, since the curves
  are folded on the refined orbital period determined in Sect.~\ref{sec:lc-superperiodicity}.}
\begin{tabular}{lccc}
\hline \noalign {\smallskip}
\hline \noalign {\smallskip}
Obs. & 3 & 5 & 4 \\
\hline \noalign {\smallskip}
Param.   &   &   &   \\
$M$ & 18.99 $\pm$ 0.03 & 19.12 $\pm$ 0.03 & 19.07 $\pm$ 0.03  \\
$A$ & 0.15 $\pm$ 0.04 & 0.25 $\pm$ 0.04 & 0.17 $\pm$ 0.04  \\
$\phi_0$  & 0.99 $\pm$ 0.04 & 0.93 $\pm$ 0.03 & 0.11 $\pm$ 0.03 \\
\hline \noalign {\smallskip}
\rchisq (d.o.f.) & 0.84 (13) & 0.29 (13) & 0.26 (13) \\  
\noalign {\smallskip} \hline \label{tab:omfit}
\end{tabular}
\end{center}
\end{table}

\subsection{Dip recurrence interval}
\label{sec:lc-ephemeris}

The most accurate orbital ephemeris for \src\ is that of
\citet{1254:motch87apj} who derived a period of $0.163890 \pm
0.000009$~day with a time of optical minimum of JD $2,445,735.693 \pm
0.004$. This ephemeris is not accurate enough to maintain phase
between the two XMM-Newton observations where deep dips were detected
in 2001 January and 2007 January as the accumulated uncertainty on the
predicted time of $\pm$2.9~hrs is close a complete orbital cycle.
Recently, \citet{1254:barnes07mn} updated the orbital ephemeris
based on an RXTE Proportional Counter Array observation on 2004 May 26 
that includes a deep dip. They estimate a new value of JD 2,453,151.509(3)
for the optical minimum ($\phi = 0$) assuming that the centre of the X-ray dip
occurred at $\phi =  0.84$.  
The centre of the first deep dip observed by XMM-Newton in 2007
January 14 (Obs~4) occurred at 3.72~hrs UTC (see Sect.~\ref{sec:x-lc}). Using the XMM-Newton dip to define
the dip phases and using the \citet{1254:motch87apj} period
gives a predicted time of 3.30~hrs for the
dip observed by RXTE, only 0.23~hrs before the dip time given 
by \citet{1254:barnes07mn}. Given that the predicted
uncertainty in the dip times is $\pm$1.3~hr, this is an 
unusually close prediction. Similarly, the predicted time for the 
dip observed by XMM-Newton in the 2001 January 22 observation (Obs~1)
is 19.10~hr, only 0.12~hrs before the time given by \citet{1254:boirin03aa}, whereas the uncertainty in the \citet{1254:motch87apj}
period determination corresponds to $\pm$2.9~hrs. 
It is unlikely that both of these close predictions
are coincidences and instead we suggest that the uncertainty
on the \citet{1254:motch87apj} ephemeris has been
overestimated by a factor of $\sim$5. This would imply a dip 
recurrence period of $0.163890 \pm 0.000002$~day.

\subsection{Light curves from other X-ray missions}
\label{sec:lc-superperiodicity}

We have shown that at the 1 or 4~s time resolution of
Fig.~\ref{fig:detail-lightcurves}, shallow dips include a few
very short duration deep dips each with depths and hardness ratios similar
to the longer duration dips seen during Obs~1 and 4 (see
Sect.~\ref{sec:x-lc}). In order to investigate the dip variability
further, we extracted from the High Energy
Astrophysics Science Archive Research Center (HEASARC) light
curves from previous \src\ X-ray observations 
with the maximum time resolution available. 
We then searched for the signature of
shallow dips that may have been missed due to
the use of too coarse binning. Our goal was to determine the minimum time
between the disappearance and re-appearance of deep and
shallow dips and thus to see if there is a characteristic time scale for these changes
which might provide a clue as to the underlying mechanism. 
An overview of the dip states seen in the different X-ray 
observations of \src, following the definition of Table~\ref{tab:obslog}, 
is given in Table~\ref{tab:dips}. 

The variations in dip depth could result from some irregular process
and occur on a range of time scales. Alternatively, there could be an
underlying regular mechanism such as disc precession which could give
rise to a characteristic time scale for the variations. In order to
investigate possible characteristic time scales we examined all
available X-ray light curves.  The longest uninterrupted interval with
deep dipping activity occurs within the 19~hr duration EXOSAT
observation on 6--7 August 1984, when five consecutive deep dips are
observed. Similarly, the longest interval without dipping activity is
observed in the RXTE pointed observation on 6--7 December 2001, when
dipping was absent during 6 consecutive cycles, or 19~hr. In
XMM-Newton Obs~5 four consecutive shallow dips are observed during
17~hr.  These measurements give a lower limit of $\approxgt$20~hr for
any characteristic time scale of a given dipping state.
In contrast, a transition from deep dipping to non-dipping is observed
during the EXOSAT observation of 5 February 1984 in only one orbital
cycle, or 4~hr. The long 60~hr (but non-continuous) RXTE pointed
observation on 9--12 May 2001 provides additional constraints. Here we
find a complex sequence of dip behaviour consisting of a deep dip, a
gap of 5 cycles (due to incomplete coverage), two non-dip episodes, a
gap of one orbital cycle, a shallow dip, a gap of another orbital
cycle, a dip that could be shallow or deep, a gap of 2 cycles, and 2
further dips that are not fully covered and could be deep or shallow.
Thus, if any of the last three dip episodes is deep, this would imply
a re-appearance of deep dipping within $\sim$60~hr.  Given that a
sharp variation in dip properties was observed within a single 4~hr
orbital cycle, that the dip patterns can be unchanged for $\sim$20~hr
intervals and that a 60~hr observation revealed the whole range of dip
depths suggests that if there is a characteristic timescale for dip 
depth variability it is around 60~hr.

\begin{table*}
\begin{center}
\caption[]{X-ray observations of \src\ dip states. Dip states are
classified according to the integrated reductions in the X-ray light
curves using the definition in Table~\ref{tab:obslog}. We have chosen
the lowest energy range available for the light curves, where the dips
are likely to become more apparent: 1--3.8~keV
(EXOSAT), 2--6~keV ($Ginga$), 2--4~keV (RXTE), 0.7--2.5~keV
($BeppoSAX$), 0.6--2~keV (XMM-Newton), and 1--8~keV ($Chandra$).}
\begin{tabular}{lclc}
\hline \noalign {\smallskip}
Date & Mission & Reference & Dip Depth \\
\hline \noalign {\smallskip}
1984 February  5          & EXOSAT & \citet{1254:courvoisier86apj}  & Shallow, Undetected\\
1984 May 15               & EXOSAT & \citet{1254:courvoisier84conf} & Shallow \\
1984 August 6--7          & EXOSAT & \citet{1254:courvoisier86apj}  & Deep \\
1985 April 14--15         & EXOSAT & \citet{1254:courvoisier86apj}  & Undetected, shallow\\
1990 August 1--3          & $Ginga$ & \citet{1254:uno97pasj}        & Deep\\
1997 April 28 -- May 1    & RXTE & \citet{1254:smalewachter99apj}   & Undetected, shallow \\
1998 December 22--23     & $BeppoSAX$ & \citet{1254:iaria01apj}     & Undetected \\
2001 January 22          & XMM-Newton & \citet{1254:boirin03aa}     & Deep \\
2001 May 9--12           & RXTE  & \citet{1254:smale02apj} & Deep, undetected, shallow \\
2001 December 6--7       & RXTE  & \citet{1254:smale02apj} & Undetected \\
2002 February 7--8       & XMM-Newton & \citet{1254:boirin03aa} & Undetected \\
2003 October 10          & $Chandra$ & \citet{1254:iaria07aa} & Undetected \\
2004 May 26              & RXTE & \citet{1254:barnes07mn} & Deep, undetected \\
\noalign {\smallskip} \hline \label{tab:dips}
\end{tabular}
\end{center}
\end{table*}

We also examined the \src\ RXTE All-Sky Monitor (ASM) 1.5--12~keV
light curves spanning 12~years and with a time resolution of 96~s to
search for any characteristic timescale for the appearance and
disappearance of dips. We corrected the ASM light curves to the solar
system barycentre using the FTOOL task {\tt faxbary} and folded the
light curves into segments between 4 and 200 orbital periods
duration using the ephemeris and period from
Sect.~\ref{sec:x-lc}, but clear dipping activity was not detected 
due to the low count rates.

\begin{figure}[!ht]
  \resizebox{\hsize}{!}{\includegraphics[angle=0, width=3cm]{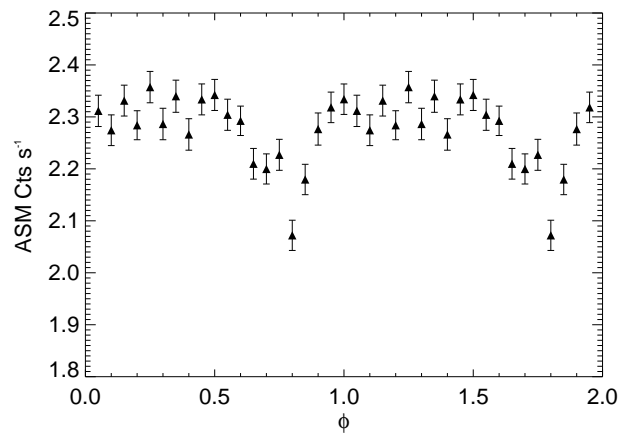}}
  \caption{ASM 1.5--12~keV background subtracted light curve for \src\ folded at the refined orbital period determined in Sect.~\ref{sec:lc-superperiodicity}, showing the clear detection of dipping at $\phi$~$\sim$~0.8. We have used the same reference phase as in Fig.~\ref{fig:omlightcurves}.}
 \label{fig:ASM-folded}
\end{figure}

Next, we investigated whether the orbital period of the
source can be refined from the ASM light curves. We first calculated
the Lomb-Scargle periodogram \citep{scargle82apj} which revealed a
clear peak at the optical period. Its probability to be due to a
random fluctuation is $\sim$ 4$\times$ 10$^{-6}$ implying a clear 
detection of the X-ray orbital modulation in the ASM data. The ASM
folded light curve shows a clear dip at the expected phase
($\phi\sim$~0.8, see Fig.~\ref{fig:ASM-folded}). The reduction in 
intensity compared to the persistent level of
2.3~s$^{-1}$ is 10\%. This low value is consistent with averaging
intervals with deep, shallow and no dipping activity. The dip is preceded by
a "shoulder" with a reduced intensity of $\sim$5\% with respect to
the persistent level at $\phi\sim$0.7. Such "shoulders" are clearly seen in the \xmm\ light curves (see Sect.~\ref{sec:x-lc}) and indicate
the high sensitivity that can be achieved using ASM light curves 
spanning a long time period. 

Although the dips may not be observed during all orbital cycles, 
they do provide a sharp fiducial mark
which can be used to determine an improved ephemeris. We used the
\chisq\ test on folded data \citep{leahy83apj} which is sensitive to
the total power, not only the fundamental, to better constrain the
value of orbital period. The optical period is well detected with a
\chisq\ of 85.10 for 21 degrees of freedom (d.o.f.). We performed Monte Carlo 
simulations in order to estimate the uncertainty on the best period.
The actual ASM time series folded at the test frequency provided a
mean pattern from which a fake light curve was created by adding
Gaussian noise with $\sigma$ proportional to the error associated to each
actual measurement. From a total of 1000 runs we derived the
distribution of the frequency at which the maximum \chisq\ of the
periodogram occurs. We find a best period of 0.16388875 $\pm$ 0.00000017~d,
consistent with that derived by \citet{1254:motch87apj} from optical data
and with the one determined in Sect.~\ref{sec:lc-ephemeris}, improving by 
a factor 50 in accuracy. We provide a revised ephemeris of 
JD~2,454,114.6877975 for the optical minimum. This ephemeris has been 
derived taking into account the dip reference phase calculated in 
Sect.~\ref{sec:x-lc} and assuming that the dip centre occurs at $\phi$ = 0.8,
and is the same used to fold the ASM light curve shown in 
Fig.~\ref{fig:ASM-folded}.

\begin{figure*}[!ht]
\includegraphics[angle=0,width=0.49\textwidth]{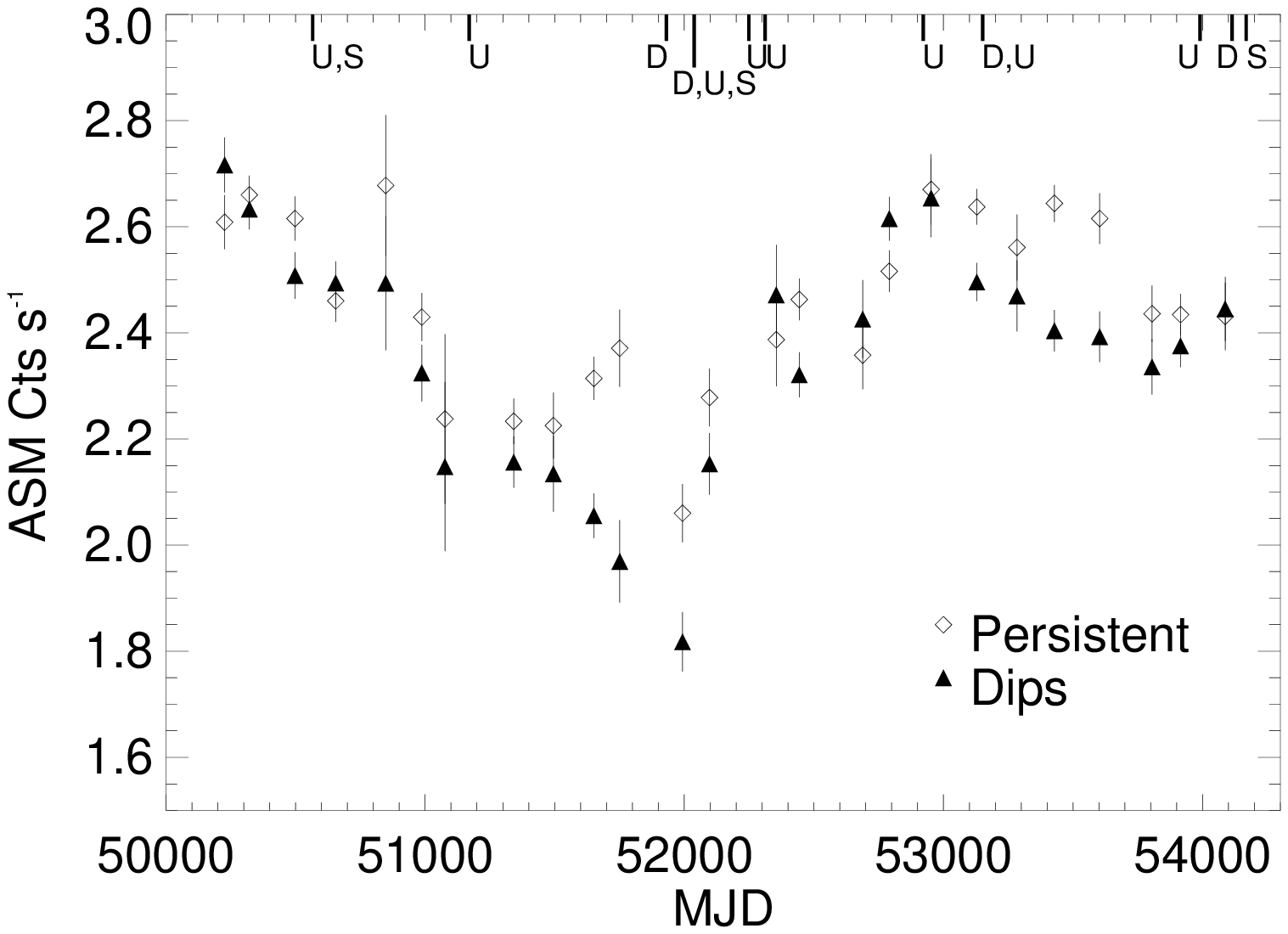}
\includegraphics[angle=0,width=0.49\textwidth]{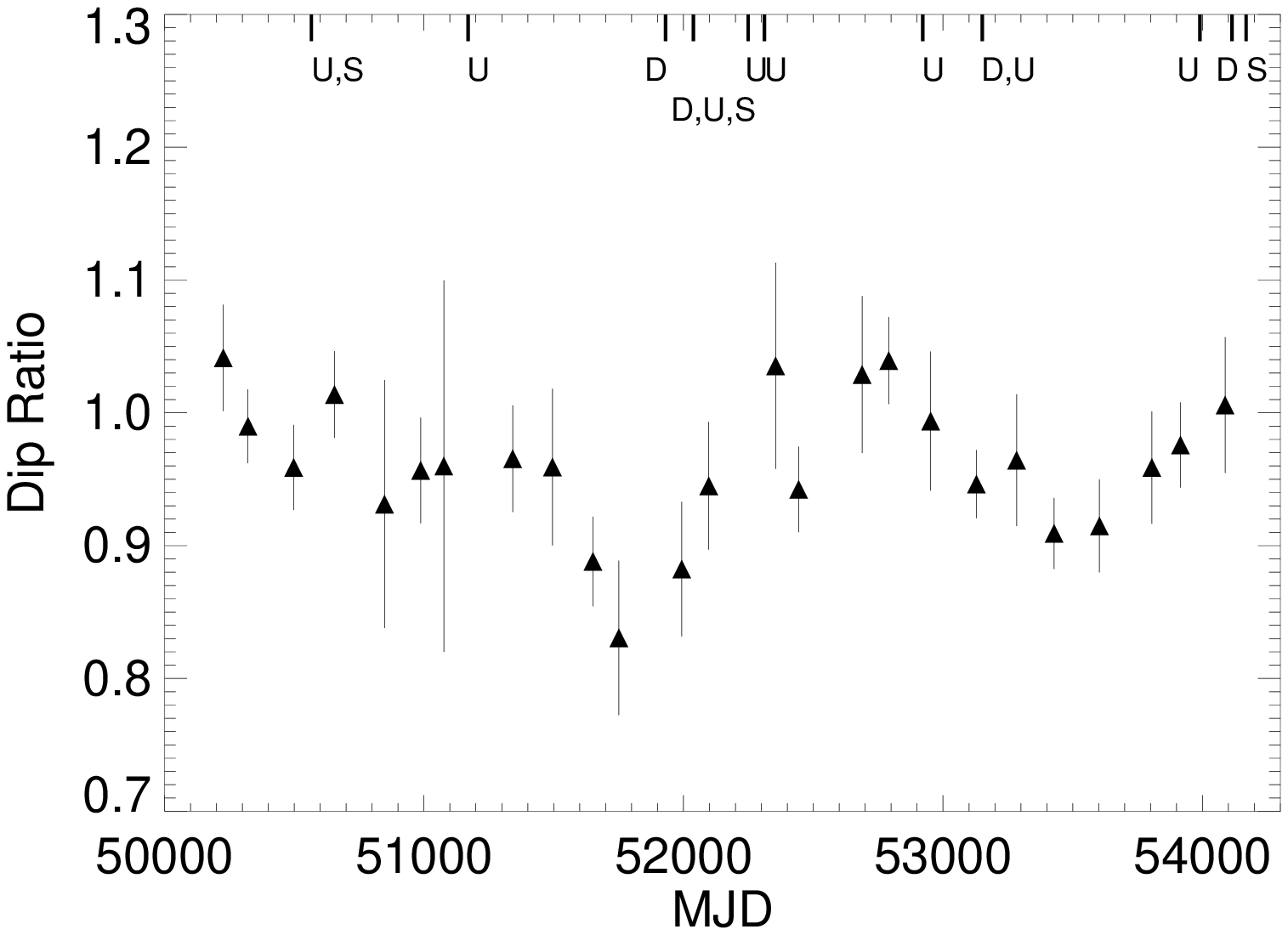}
\caption{ASM 1.5--12~keV background subtracted light curve for \src\ with a binning of 1000 orbital periods or 163.9 days during persistent phase ($\phi$ = 0.0--0.6, open diamonds) and dipping phase ($\phi$ = 0.6--1.0, filled triangles) ({\it left panel}) and the ratio of dipping to persistent phase ({\it right panel}). The thick vertical tick marks indicate the dipping state of the source during pointed observations (U: Undetected, S: Shallow, D: Deep; see Table~\ref{tab:dips}). The increase in dip depth around MJD~52000 may be associated with a reduction in persistent count rate.}
 \label{fig:asm-lc-1000period-alldata.ps}
\end{figure*}

Finally, we studied the long-term dipping behaviour of \src\ with the available ASM data. Fig.~\ref{fig:asm-lc-1000period-alldata.ps} shows the count rate during dipping ($\phi$ = 0.6--1.0) and persistent phases ($\phi$ = 0.0--0.6) (left panel) and the ratio of these quantities (right panel) along the 12~years of ASM data. In general, the periods of low persistent count rate ($\approxlt$2.2 cts s$^{-1}$) show a ``Dip Ratio'' significantly below 1, indicating strong dipping activity. In contrast, the pointed observations showing deep dipping activity (see Table~\ref{tab:dips}) are distributed along the whole time range. 
Given the small amount of coverage of pointed observations, we cannot exclude a relation between the detection of deep dipping in pointed observations and the ASM count rate.

\subsection{X-ray spectra}
\label{sec:spectra}

For the observations showing deep and shallow dipping, we first selected
intervals of "persistent" or dip free emission. Spectra were
accumulated corresponding to these intervals and for the entire dip
free observations. We rebinned the EPIC pn spectra to over-sample the
$FWHM$ of the energy resolution by a factor 3 and to have a minimum of
25 counts per bin, to allow the use of the $\chi^2$ statistic. We
rebinned the RGS spectra to over-sample the $FWHM$ of the energy
resolution by a factor 2 to be sensitive to narrow features and
we used the C-statistic. We did not rebin the JEM-X and ISGRI spectra
to avoid losing spectral resolution. We performed spectral analysis
using XSPEC \citep{arnaud96conf} version 12.3.1. We used the
photo-electric cross-sections of \citet{wilms00apj} to account for
absorption by neutral gas with solar abundances (the {\tt tbabs} XSPEC
model). Spectral uncertainties are given at 90\% confidence
($\Delta$\chisq = 2.71 for one interesting parameter), and upper
limits at 95\% confidence.

\subsubsection{EPIC pn and INTEGRAL spectral analysis}
\label{subsec:pn-isgri}

We fit the 0.6--10~keV EPIC pn, 5--25~keV JEM-X and 15--70~keV ISGRI
spectra of \src\ with a model consisting of a disc blackbody and a
thermal comptonisation model, {\tt comptt}, both modified by
photo-electric absorption from neutral and ionised material together
with two Gaussian emission features at $\sim$1.0~keV and $\sim$6.4~keV
modified by neutral absorption only and one Gaussian absorption
feature at $\sim$1.8~keV ({\tt tbabs*warmabs*(diskbb+comptt) +
tbabs*(gau+gau) +gau)}. The feature at 1.8~keV is probably due to an
incorrect modeling of the Si absorption in the CCD detectors by the
EPIC pn calibration and is therefore not further discussed. Constant
factors, fixed to 1 for the EPIC pn spectrum, but allowed to vary for
the JEM-X and ISGRI spectra, were included multiplicatively in order
to account for cross-calibration normalisation uncertainties. The {\tt
warmabs} component models the absorption due to a photo-ionised plasma
in the line of sight. All relevant ions are automatically taken
into account, including those having small cross-sections, which can
contribute significantly to the absorption when combined. The relative
column densities of the ions are coupled through a photo-ionisation
model.  During the fitting process, {\tt warmabs} calculates spectra
using stored level populations pre-calculated with XSTAR
\citep{kallman01apjs} which are then scaled using abundances specified
during the XSPEC session before the spectra are calculated. A given
ionising continuum spectrum is assumed when calculating the level
populations with XSTAR (see below for details on the ionising
continuum used in this paper). All the models of XSTAR assume a
spherical gas cloud with a point source of continuum radiation at the
centre. Therefore it implicitly assumes spherical symmetry and
radially beamed incident radiation. The main parameters of {\tt
warmabs} are \nhwarmabs, $\xi$, \sigmav, and $v$, representing the
column density of the absorber, the ionisation parameter, the
turbulent velocity broadening, and the average systematic velocity
shift of the absorber (negative values indicate blueshifts). The
ionisation parameter is defined as $\xi=L/nR^2$, where $L$ is the
(energy) luminosity of the incident radiation integrated from 1 to
1000 Ry, $n$ is the gas density, and $R$ is the distance from the
radiation source. The {\tt warmabs} component is necessary to
account for the complex residuals evident near 7~keV as well as for
modifying the overall continuum shape.

After performing fits for all the
observations, it was evident that the disc-blackbody component was
changing significantly among observations while the thermal
comptonisation component was unchanged within the
uncertainties. 
Therefore, we fit the spectra for all observations
together tying the parameters of the thermal comptonisation, except the
normalisation factor, and Gaussian components for all the observations. 
We kept the normalisation factor of the comptonisation component untied 
to prevent an artificial variation of the disc-blackbody flux due to the 
constancy of the former component.  
When fitting the {\tt warmabs} component, we first searched for the 
lowest possible value of
the ionisation parameter, $\xi$, which fits well the absorption lines,
and then we fixed the parameter to this value to obtain the final fits
and the uncertainties. This is done to prevent a fit with ``artificially''
increased continuum normalisations and unrealistic column densities
for the absorber.

\begin{table*}
\begin{center}
\caption[]{Best-fits to the 0.6--10~keV EPIC pn, 5--25~keV JEM-X and
  15--70~keV ISGRI persistent (non-dip) spectra for all the
  observations using the {\tt tbabs*warmabs*(dbb+comptt)+
  tbabs*(gau$_1$+gau$_2$) model.} \kcomptt, \kbb\ and \kgau\ are the
  normalizations of the comptonisation component, disc blackbody and
  Gaussian emission features, respectively. $kT_0$ is the input soft  
  photon (Wien) temperature, \ktcomptt\ the plasma temperature, and 
  $\tau_{p}$ the plasma optical depth, of the comptonisation component.
  \ktbb\ is the temperature of the disc blackbody. \egau\ and $\sigma$ 
  represent the energy and width of the Gaussian features. \nhabs\ and
  \nhwarmabs\ are the column densities for the neutral and ionised 
  absorbers, respectively. $\xi$, \sigmav, and $v$ are the ionisation
  parameter (in units of erg cm s$^{-1}$), the turbulent velocity 
  broadening, and the average systematic velocity shift of the absorber
  (negative values indicate blueshifts). $F$ is the unabsorbed
  0.6--10 keV total flux and $F_{dbb}$ is the unabsorbed 0.6--10 keV
  disc blackbody flux. The widths ($\sigma$) of the Gaussian emission
  lines {\tt gau$_{1-2}$} are constrained to be $\le$1~keV and
  $\le$0.1~keV, respectively. \nhabs\ is linked for Obs~1--2 and for 
  Obs~3--5 during the fits.}
\begin{tabular}{lcccccc}
\hline \hline\noalign{\smallskip}
Observation No. & & 2 & 3 & 5 & 1 & 4 \\
Dips            & & Undetected & Undetected & Shallow & Deep & Deep \\ 
\noalign{\smallskip\hrule\smallskip}
& Comp. & & & & &    \\
Parameter & & & &  &  & \\
& {\tt comptt} & & & &  & \\
$kT_0$ (keV) & & \multicolumn{5}{c}{0.173 $\pm$ 0.001} \\
\multicolumn{2}{l}{\ktcomptt\ (keV)} & \multicolumn{5}{c}{3.06 $\pm$ 0.03}\\
\multicolumn{2}{l}{$\tau_{p}$} & \multicolumn{5}{c}{6.07 $\pm$ 0.05}\\
\multicolumn{2}{l}{\kcomptt\ } & 0.0747 $\pm$ 0.0005 & 0.0772 $^{+0.0003}_{-0.0015}$ & 0.0763 $^{+0.0006}_{-0.0003}$ & 0.0740 $\pm$ 0.0008 & 0.0748 $\pm$ 0.0006 \\
& {\tt dbb} & & &  & &\\
\ktbb\ {\small(keV)} & & 2.128 $\pm$ 0.006 & 2.069 $\pm$ 0.004 & 1.86 $\pm$ 0.01 & 1.66 $\pm$ 0.01 & 1.74 $\pm$ 0.01 \\
\multicolumn{2}{l}{\kbb\ {\small[(R$_{in}$/D$_{10}$)$^{2}$ cos$\theta$]}} & 1.10 $\pm$ 0.01 & 1.363 $\pm$ 0.009 & 1.76 $^{+0.01}_{-0.03}$ & 2.34 $\pm$ 0.05 & 2.15 $\pm$ 0.04 \\
& {\tt gau$_1$}  & & & & & \\  
\multicolumn{2}{l}{ \egau\ {\small(keV)}} &  \multicolumn{5}{c}{6.59 $\pm$ 0.05}  \\
\multicolumn{2}{l}{ $\sigma$ {\small(keV)}} & \multicolumn{5}{c}{0.53 $\pm$ 0.06} \\
\multicolumn{2}{l}{ \kgau\ {\small(10$^{-4}$ ph cm$^{-2}$ s$^{-1}$)}} & \multicolumn{5}{c}{3.3 $\pm$ 0.3} \\
& {\tt gau$_2$}  & & & & & \\  
\multicolumn{2}{l}{ \egau\ {\small(keV)}} & \multicolumn{5}{c}{1.046 $\pm$ 0.004} \\
\multicolumn{2}{l}{ $\sigma$ {\small(keV)}} & \multicolumn{5}{c}{0.1}  \\
\multicolumn{2}{l}{ \kgau\ {\small(10$^{-3}$ ph cm$^{-2}$ s$^{-1}$)}} & \multicolumn{5}{c}{1.69 $^{+0.04}_{-0.07}$ }\\
& {\tt tbabs} & & &   \\
\multicolumn{2}{l}{\nhabs\ {\small($10^{22}$ cm$^{-2}$)}} & 0.225 $^{+0.007}_{-0.004}$ & 0.220 $\pm$ 0.001 & 0.220 $\pm$ 0.001 & 0.225 $^{+0.007}_{-0.004}$ & 0.220 $\pm$ 0.001 \\
& {\tt warmabs} & & & &  & \\
\multicolumn{2}{l}{\nhwarmabs\ {\small($10^{22}$ cm$^{-2}$)}} & 2.2 $\pm$ 0.4 & 2.3 $\pm$ 0.3 & 2.0 $\pm$ 0.3 & 2.8 $\pm$ 0.5 & 2.6 $\pm$ 0.3 \\
\logxi\ {\small(\xiunit)} & & 3.95 (f) & 3.95 (f) & 3.95 (f) & 3.95 (f) & 3.95 (f) \\
\sigmav\ {\small(km s$^{-1}$)} & & 4400 $^{+2300}_{-1900}$ & 3650 $\pm$ 1000 & 2600 $^{+1700}_{-1300}$ & 4400 $^{+3100}_{-1900}$ & 1500 $^{+700}_{-600}$  \\
$v$ {\small(km s$^{-1}$)} & & -240 $\pm$ 1200 & -2040 $\pm$ 600 & -1440 $\pm$ 780 & 240 $^{+1500}_{-900}$ & -1380 $\pm$ 360 \\ 
\noalign {\smallskip}
\noalign {\smallskip}
\hline\noalign {\smallskip}
\multicolumn{2}{l}{\rchisq (d.o.f.)} & \multicolumn{5}{c}{1.51 (1147)}\\
\hline\noalign {\smallskip}
\hline\noalign {\smallskip}
\multicolumn{2}{l}{$F$ \small (10$^{-10}$ \ergcmsec)}  & 8.9 & 9.5 & 8.8 & 8.0 & 8.4 \\
\multicolumn{2}{l}{$F_{dbb}$ \small (10$^{-10}$ \ergcmsec)} & 4.3 & 4.7 & 4.1 & 3.4 & 3.8 \\
        \multicolumn{2}{l}{Exposure (ks)} & 15.0 & 43.3 & 26.6 & 9.8 & 47.0 \\
\noalign{\smallskip\hrule\smallskip}
\label{tab:bestfit}
\end{tabular}
\end{center}
\end{table*}

\begin{figure*}[!ht]
\includegraphics[angle=0,width=0.55\textwidth]{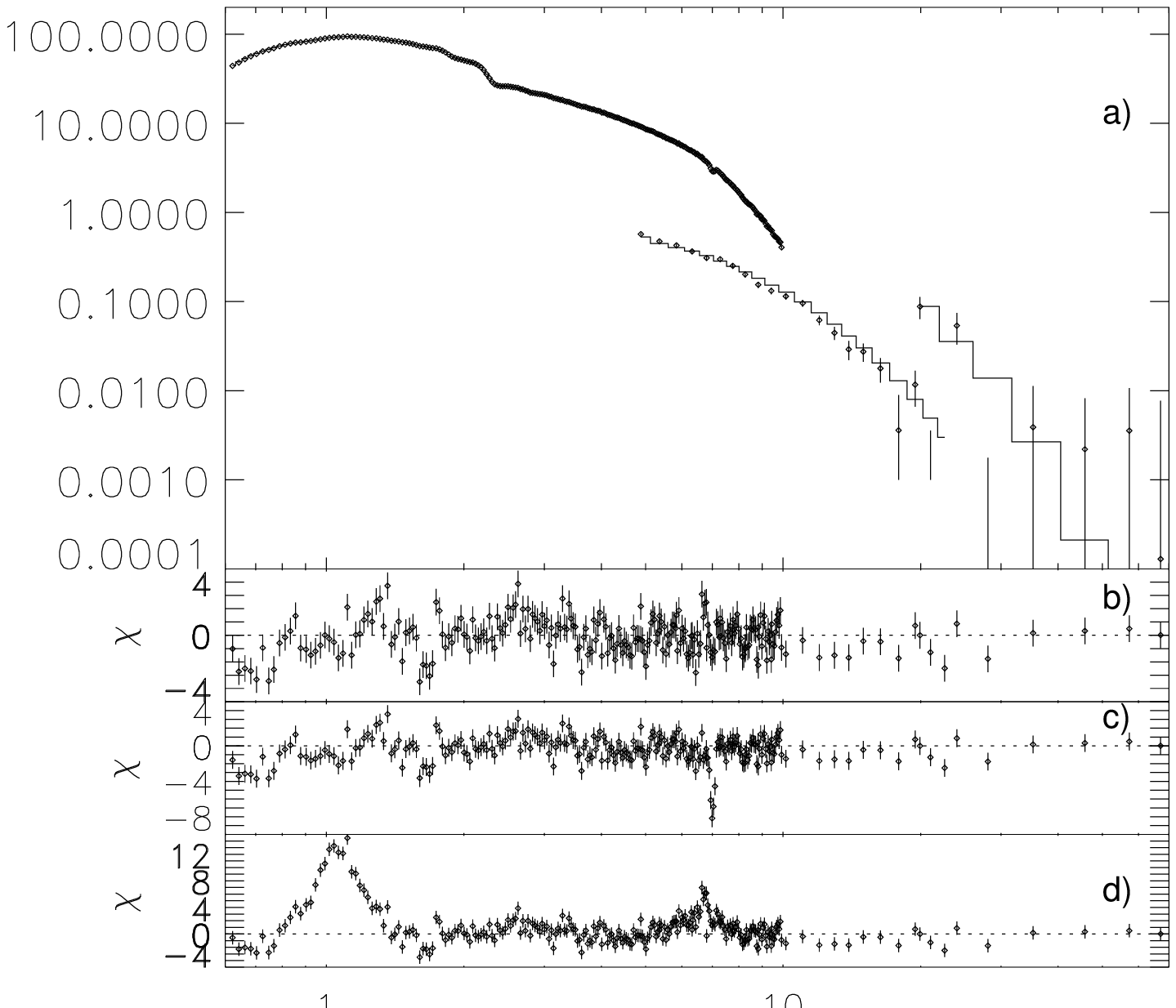}
\hspace{-0.5cm}
\includegraphics[angle=0,width=0.55\textwidth]{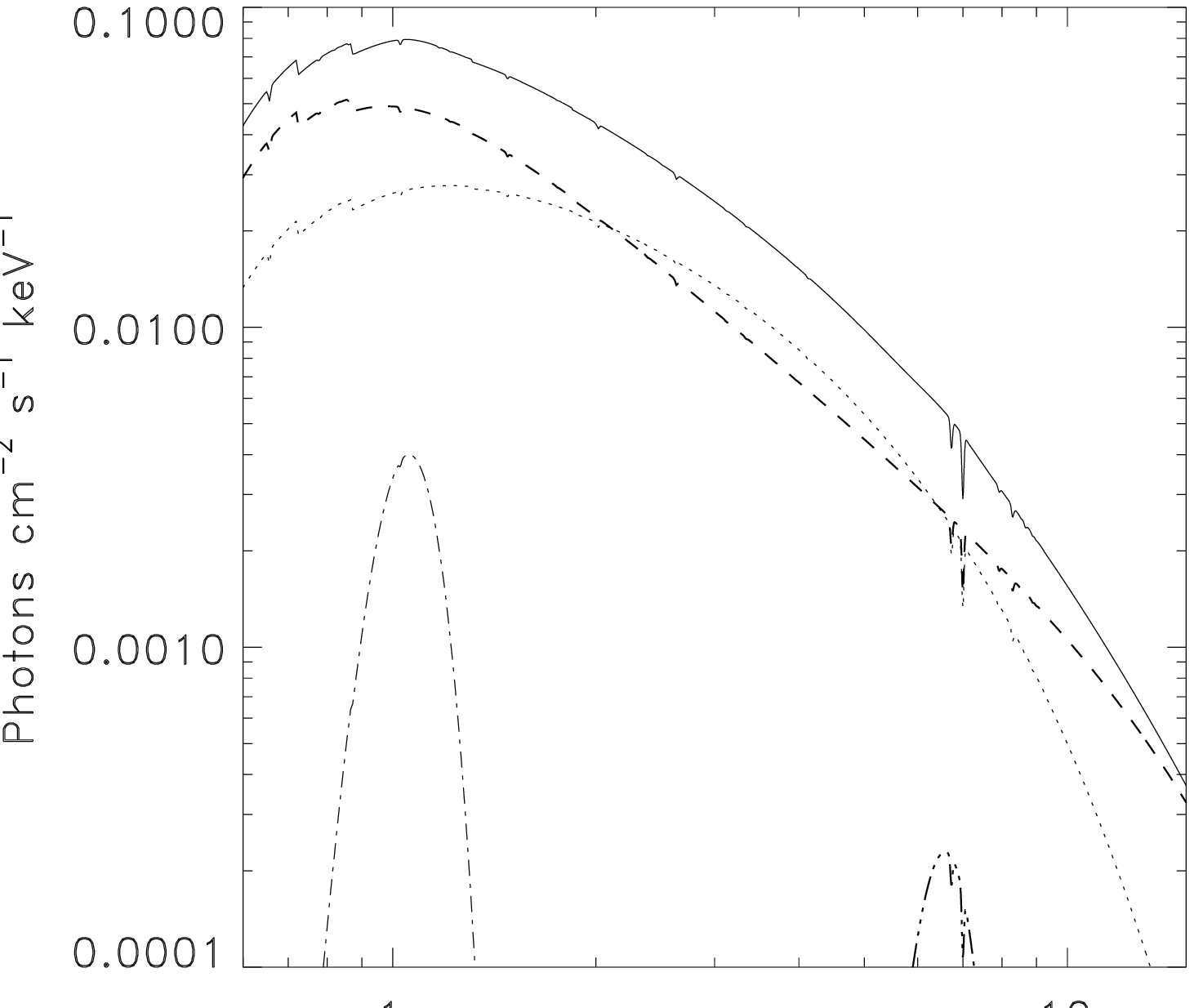}
\vspace{0.5cm}
\caption{{\it Left}: 0.6--10 keV EPIC pn, 5--25~keV JEM-X and 15--70~keV ISGRI
\src\ Obs~4 spectra fit with a disc-blackbody ({\tt diskbb}) and a
thermal comptonisation component ({\tt comptt}) as continuum modified
by absorption from neutral ({\tt tbabs}) and ionised ({\tt warmabs})
material together with two Gaussian emission lines ({\tt gau$_{1-2}$})
which are modified by absorption from neutral material ({\tt tbabs})
only. (b) Residuals in units of standard deviation from the above
model. (c) Residuals when \nhwarmabs\ is set to 0 showing the
$\sim$7~keV absorption features clearly.  (d) Residuals when the
normalisations of the Gaussian emission lines are set to 0. Best-fit
parameters are given in Table~\ref{tab:bestfit}. {\it Right}: Best-fit
model. Solid, dashed, dotted, dashed-dotted and dashed-dotted-dotted
lines represent, respectively, the total model, the comptonisation
component, the disc-blackbody component and the Gaussian features at 1
and 6.6~keV.
\label{fig:spec}}
\end{figure*}

The \rchisq\ of the resulting fit is 1.51 for 1147 d.o.f.. 
The parameters of the best-fit model are given in
Table~\ref{tab:bestfit} and the residuals of the fit are shown in
Fig.~\ref{fig:spec}. The values obtained for the neutral absorption
are significantly lower than the neutral absorption predicted in the
direction of \src\ by \citet{nh:dickey90araa}, $0.31\times 10^{22}$
cm$^{-2}$, and are closer to the value reported in
\citet{nh:kalberla05aa} of $0.22\times 10^{22}$ cm$^{-2}$.

Substituting the thermal comptonisation component by a cutoff
power-law component increases the \rchisq\ of the fit to \rchisq\ =
2.18 for 1150 d.o.f.. The parameters of the fit remain the same as in
Table~\ref{tab:bestfit} except the neutral absorption, which increases
to \nhabs\ = $(0.307 \pm 0.001)\times 10^{22}$ cm$^{-2}$ (Obs~1--2)
and $(0.298 \pm 0.001)\times 10^{22}$ cm$^{-2}$ (Obs~3--5). The
temperatures (normalisations) of the disc-blackbody component are
systematically higher (lower) for all observations compared to the
best-fit model, but the relative differences among observations remain
the same. The photon index, $\Gamma$, and cutoff energy of the power-law are
2.08 and 34~keV, respectively. The value of the energy cutoff is
significantly larger than the one obtained from fits to RXTE data by
\citet{1254:smale02apj}. Substituting the thermal comptonisation
component by a power-law component also increases the \rchisq\ of
the fit to \rchisq\ = 2.14 for 1151 d.o.f.. The neutral absorption
increases to \nhabs\ = $(0.321 \pm 0.001)\times 10^{22}$ cm$^{-2}$
(Obs~1--2) and $(0.310 \pm 0.001)\times 10^{22}$ cm$^{-2}$
(Obs~3--5). The index, $\Gamma$, of the power-law is 2.25.

Substituting the disc-blackbody component by a blackbody component
also increases the \rchisq\ of the fit to \rchisq\ = 1.90 for 1133
d.o.f.. However, in order to get an acceptable fit the
thermal comptonisation component and the neutral absorption must be
allowed to vary for each observation.

We used the best-fit comptonisation component as an ionising continuum
for the photoionised absorber. In a previous paper
\citet{ionabs:diaz06aa} used a different ionisation continuum,
consisting in a power law with $\Gamma$ = 1.28 (obtained from a fit to
the Obs~1 EPIC pn spectrum with only a power law) and a cutoff at
5.9~keV (as obtained from fits to RXTE data by
\citet{1254:smale02apj}). However, the broad band XMM-Newton and
INTEGRAL spectrum available for Obs~4 is better fit with a continuum
consisting of a disc blackbody and a thermal comptonisation component
than with the model used by \citet{1254:smale02apj}. In order to 
evaluate the influence of the ionisation continuum in our fits, we 
generated two ionisation continua: one with the two main components of the continuum
(disc-blackbody and comptonisation component) and another one with
only the comptonisation component. The warm absorber is very strongly
ionised with only absorption features from H-like \fetsix\
evident. Thus the continuum responsible for ionising the absorber must
be the high-energy component, which is in this case the
comptonisation component. Including the disc-blackbody component in
the ionising continua did not change the value of the warm absorber
ionisation parameter significantly. Thus, we only included the comptonisation
component in the ionising continuum.

\subsubsection{RGS spectral analysis}

\begin{figure}[!ht]
  \resizebox{\hsize}{!}{\includegraphics[angle=0, width=3cm]{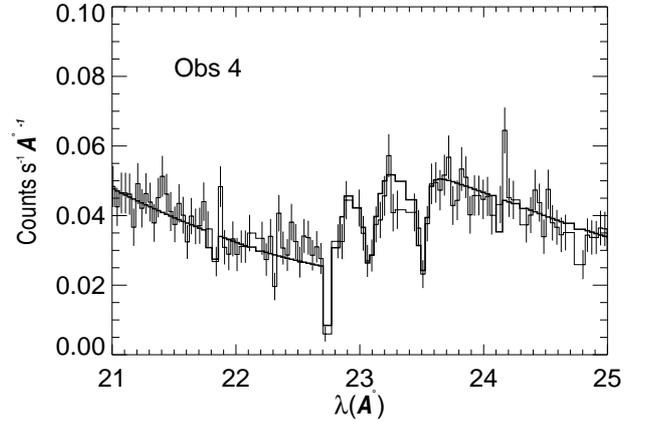}}
  \caption{\src\ 21-25~\ang\ RGS spectra for Obs~4, which has the
longest exposure. The structure around the O edge with clear
absorption features at the 1s--2p transitions of \oone\ and \othree\ is shown.}
 \label{fig:rgsspectrum}
\end{figure}

We examined the 0.3--2~keV (6--38~\ang) first and second order
RGS spectra. A joint analysis of RGS and pn data was not possible due
to the presence of strong emission features in the pn below 3~keV
which were not detected in the overlapping energy range of the RGS
(see Sect.~\ref{subsec:pn-isgri}).
We could fit the RGS spectra of all the
observations with a continuum consisting of a power law modified by
photo-electric absorption from neutral material (C-statistic between 
3200 and 4000 for $\sim$2820 d.o.f. for Obs~1 to 5). The RGS
spectra show structured residuals near the O edge at 0.54~keV
(see Fig.~\ref{fig:rgsspectrum}). Around this energy there 
is a $\sim$25\% change in the
instrument efficiency, which may not be fully accounted for in the
data processing. Thus, the residuals could be due either to a
calibration uncertainties or have an astrophysical origin.

We investigated the second possibility by studying in detail the
region around 0.54~keV (23~\ang). We searched for the signature of O
absorption in the interstellar medium similar to that observed by
\citet{cygx2:takei02apj} and \citet{cygx2:costantini05aa} from
Cyg\,X-2 and by \citet{juett04apj} from a number of galactic
sources. We used the {\tt tbvarabs} absorption model from
\citet{wilms00apj} with the oxygen abundance set to zero to allow for
an explicit modeling of the oxygen absorption. We included narrow
Gaussian absorption features at energies of 0.527 and 0.537~keV (23.5
and 23.1~\ang, respectively) and an edge at 0.543~keV (22.8~\ang).
This improves the fit significantly and we identify the common feature
at 0.527~keV with absorption from \oone\ 1s--2p and the feature at
0.537~keV with absorption from \othree\ 1s--2p. We note that the \otwo\
1s--2p absorption feature expected at $\sim$0.531~keV (23.3\ang) falls
into an RGS bad column and cannot therefore be detected. In addition,
an absorption feature at 0.711~keV (17.4~\ang) is evident in the
first and second order RGS spectra of all the observations. We
attribute this feature to absorption by the L3 (2p$_{3/2}$) edge of
interstellar Fe.  Including narrow Gaussian absorption features to
model these residuals reduces the C-statistic to 3000--3400 for 2800
d.o.f. for Obs~1 to 5.

None of the observations show any additional narrow features and including
absorption from a photo-ionised plasma in the model does not improve
the fit quality. The upper limits to the equivalent widths of highly
ionised features such as \neten\ and \mgtwelve\ are 0.35~eV and
0.23~eV (Obs~3), respectively.  This result is consistent with the
equivalent widths of the \neten\ and \mgtwelve, 0.13 and 0.20~eV,
respectively, predicted for an ionised absorber of log $\xi $ = 3.95,
as observed in the EPIC pn exposures.

\section{Discussion}

We have analysed five XMM-Newton observations of the LMXB \src. The
0.6--10~keV EPIC pn lightcurves are highly variable with two
observations showing deep dips, one showing shallow dips and two
without evidence for dipping behaviour. We show that the dips
identified as "shallow" when plotted with a time resolution of 120~s,
include a small number of rapid, deep, variations, when 
plotted with a time resolution of 1~s. The properties 
of these features appear similar to those
observed during deep dipping episodes, where many more such rapid
variations can combine forming the much longer deep dipping intervals
observed. The X-ray continua of the dip-free intervals are well fit
by a model consisting of a disc-blackbody with a thermal
comptonisation component extending to higher energies. Both components
are absorbed by neutral and ionised material. The temperature of the
disc blackbody is lower for observations with deep dips (1.66 $\pm$
0.01 and 1.74 $\pm$ 0.01~keV), increases for the observations with
shallow dips (1.86 $\pm$ 0.01~keV) and is highest for the observations
where dips are absent (2.128 $\pm$ 0.006 and 2.069 $\pm$ 0.004~keV). 
Because the normalisation and temperature of the disk blackbody
are highly correlated, it is necessary to examine the confidence
contours to evaluate the error range properly.
We calculated the confidence contours and found that
the range of uncertainty for the temperature is at most $\pm$0.05 keV.
This means that the change of the temperature is highly significant
even if the parameter correlation is taken into account.
We
did not find any significant changes in the ionisation parameter of
the absorber between the persistent emission from the different
observations. However, the absorber is very strongly ionised and only
features from H-like \fetsix\ are evident, making it more difficult to
detect ionisation changes compared to a less ionised plasma where
spectral features from a range of ions are visible. In addition, we
have taken advantage of the optical monitoring afforded by the OM on
XMM-Newton to produce simultaneous folded optical and X-ray light
curves of the three long observations which (luckily) cover, shallow,
deep and dip free intervals. Finally, we searched for a
characteristic dip appear-disappear timescale in the available \src\
X-ray light curves. The RXTE ASM light curves, spanning over twelve
years, are highly suited for detecting long term variability, which in
the case of \src, we expect to be of the order of few days. We have
explicitly looked for a recurrence period in the appearance of dips in
the light curves. Though we find indications for a characteristic
timescale of $\sim$60~hr in the RXTE PCA pointed observations, the RXTE ASM
light curves are not sensitive enough to confirm such timescale. 
Clearly, 
further observations are necessary. 
The RXTE ASM curves show however a period with a persistent count rate
$\approxlt$2.2 cts s$^{-1}$ spanning $\sim$~2~years, which is associated 
with deep dipping activity.
Therefore, there seems to be two processes with different timescales into 
play which determine the appearance and disappearance of dipping activity.

\subsection{Light curve modelling}

The folded OM light curve for Obs~3 is similar to those obtained by
\citet{1254:smalewachter99apj}. We find the same optical average
magnitude, V=19.0, and amplitude, $\Delta$V = 0.3, as in the 1997
optical observations. Both in the 1997 RXTE observation and in Obs~3 X-ray
dips are absent. In Obs~4, the folded lightcurves are very similar to
the February 1984 observations shown in
\citet{1254:courvoisier86apj}. They report an average magnitude V
= 19.0 and $\Delta$V = 0.4 for the optical modulation. However, we
extrapolate from their Fig.~3 an average magnitude of V = 19.1. These
values are consistent with those shown in Table~\ref{tab:omfit} for
Obs~4. Finally, whilst the RXTE light curve from
\citet{1254:barnes07mn} shows one deep dip, the shape of the
simultaneous optical curve seems to be more similar to that of \xmm\
Obs~5. Unfortunately, Fig.~3 of \citet{1254:barnes07mn} is in units of
flux density rather than magnitude and we cannot easily compare their
results with our fits. However, we note that the optical curve from
\citet{1254:barnes07mn} covers several cycles while only one X-ray
cycle is shown. We have observed that the X-ray behaviour has changed
in the cycles following the dip (see Table~\ref{tab:dips}), so this
may explain the shape and amplitude of Fig.~3 from
\citet{1254:barnes07mn}.

\citet{1254:smalewachter99apj} interpret the decrease in optical
magnitude and the cessation of dipping activity in the August 1984 to
the 1997 observations as an indication of a decrease in the bulge
size, implying that the bulge contributes $\sim$35--40\% to the
optical modulation. In contrast, \citet{1254:motch87apj} interpreted
the shape and phase of the optical modulation as indicating that the
majority of the optical modulation is caused by viewing varying
aspects of the X-ray heated hemisphere of the companion star with the
contribution of the material responsible for the X-ray dips being
responsible of $<$15\% of the disc emission. The \xmm\ OM light curves
show an optical modulation which is $\sim$20\% of the disc emission, 
implying that the results of \citet{1254:smalewachter99apj} 
and \citet{1254:motch87apj} are in disagreement. In the first
interpretation, the majority of the steady component from the optical
flux comes from the disc. During dipping intervals, cooler material at
the disc rim could obscure part of the hotter, inner disc regions. The
disappearance or reduction in size of the azimuthal structure could
lead to a more direct viewing of the hot inner disc regions and an
increase in the apparent mean disc emission temperature. This
interpretation explains naturally the increase in the temperature of
the disc with decreasing dip depth that we observe. However, following
this we should observe the $\sim$35--40\% bulge contribution to the
optical modulation mainly at $\phi$ = 0.8, when the X-ray dips are
observed, and even a higher contribution at $\phi$ = 0.3, since the
inner bulge should be brighter. Clearly, though we see a
significant excess of optical emission at $\phi$ = 0.8 during deep
dipping observations, compared to persistent and shallow dipping
observations (see Fig.~\ref{fig:omlightcurves-subtracted}), we do not
see such excess at $\phi$ = 0.3, and the fact that the optical minimum
is observed at $\phi \sim 0$ is in favour of the optical modulation
being caused by the heated hemisphere of the companion.

\subsection{Spectral variability}

We found an $apparent$ blueshift for the \fetsix\ absorption feature
in Obs~3--5. {\it Chandra} HETGS observations of the black hole
candidates \threethreenine, \xte\ and \seventeen\
\citep{gx339:miller04apj,1743:miller04apj} have revealed the presence
of variable, blue-shifted, highly-ionised absorption features which
are interpreted as evidence for outflows. In contrast, no blueshifts
have been detected in any of the highly-ionised absorbers present in
dipping LMXBs \citep{ionabs:diaz06aa}.  However, these results are all
obtained with the EPIC pn, which has a factor $\sim$4 poorer
resolution than the HETGS at 6~keV limiting the sensitivity to shifts
to be $\approxgt$1000~\kms. The grating observations of dippers yield
strict upper limits to the velocity shifts. We analysed archival data
of \mxb, the only LMXB which shows absorption features of \oeight\ and
\neten\ in the RGS \citep{1658:sidoli01aa}, and obtained a velocity
shift of $-$215$^{+245}_{-270}$ \kms\ for the absorber indicating only
a marginal blueshift. Similarly, an upper limit of $\sim$250 \kms\ is
obtained in {\it Chandra} HETGS observations of \nineteen\
\citep{1916:juett06apj}. The blueshifts found here for \src\ in
Obs~3--5 seem too large, compared to the maximum shifts of
$\sim$400~\kms\ obtained elsewhere in LMXBs with a neutron star
\citep[see e.g.,][]{gx13:ueda01apjl} and of $\sim$1000~km~s$^{-1}$ in
microquasars \citep[see e.g.,][]{1655:miller06nat}. Unfortunately, we
do not detect any discrete absorption features from the warm absorber
in the RGS, which has a much better energy resolution than EPIC.  The
lack of spectral features in the RGS wavelength range is consistent
with the high degree of ionisation of the absorber. Thus, a more
accurate measurement of the blueshifts is not possible for these
observations. However, \citet{1254:iaria07aa} report an energy for the
\fetsix\ absorption line of 6.962$^{+0.012}_{-0.015}$~keV from {\it
Chandra} HETGS observations of \src, implying that no significant
shift is present.  Therefore, it seems plausible that the shifts
detected with the pn camera do not have an astrophysical origin but
are instead due to calibration uncertainties, likely due to an
incorrect application of the Charge Transfer Inefficiency (CTI) correction
in the EPIC pn camera when high count rates are
present\footnote[2]{More information about the Charge Transfer
Inefficiency correction can be found in the {\it EPIC status of
calibration and data analysis}, by Kirsch et al. (2007), at
http:$\slash\slash$xmm.esac.esa.int}. The lines may be also
artificially broadened if the inaccuracy of the CTI correction is different for
photons arriving at the centre and at the edges of the Point Spread Function.

In addition, a moderately broad
($\sigma$ = 0.5~keV) Fe emission line at 6.6~keV with \ew$\sim$50~eV
and an emission line at 1~keV with \ew$\sim$12~eV, are present in the
spectra. The feature at 1~keV is detected in a number of X-ray
binaries and has been previously modelled either as an emission line,
or as an edge, and its nature is unclear
\citep[e.g.,][]{cygx2:kuulkers97aa, 1658:sidoli01aa, 1254:boirin03aa,
1916:boirin04aa, 1323:boirin05aa, ionabs:diaz06aa}. Provided the
feature has an astrophysical origin, its appearance always at similar
energy points to line emission, since a soft component due to a
e.g. blackbody emission is expected to change its temperature for
different sources. Its energy is consistent with a blend of \nenine\
and \neten\ emission, or Fe L emission. Recently, broad skewed Fe~K
emission lines have been discovered from a LMXB containing a neutron
star \citep[][]{serx1:bhattacharyya07apjl, serx1:cackett07apjl}. The
emission feature detected in the \src\ XMM-Newton observations at
6.6~keV is weak and moderately broad. It is not necessary to invoke
relativistic effects to explain the moderate line width, which could
be due to mechanisms such as Compton broadening. The line is broader
and has a larger equivalent width when the disc-blackbody continuum
component is substituted by a blackbody. However, the fit quality is
worse and the line is unrealistically large. Due to the availability
of XMM-Newton and INTEGRAL simultaneous data, we are able to
accurately determine the broad band continuum shape and thus we find
no evidence for a relativistically broadened Fe line in \src.

\subsection{Explanation for the changes in optical and X-ray emission}

We propose that the modulation of the optical light curves is due 
primarily to obscuration by a
precessing accretion disc that is tilted out of the orbital 
plane. The model is similar
to that proposed by \citet{herx1:gerend76apj} for Her X-1, and explains
naturally the appearance and disappearance of dips observed in the
X-ray emission of \src, as well as the spectral changes in the
continuum seen in different observations. In such a model, the accretion disc
performs three major roles in the optical flux variations: the manner
in which the disc's X-ray shadow cyclically changes the illumination
of the stellar companion, the periodic occultation of the companion by
the disc, and the changing optical brightness of the disc itself due
to the changing aspect. It is this varying aspect which may cause
the changes observed in X-ray luminosity and disc temperature and in
addition the periodic appearance and disappearance of dips in \src\
due to its particular inclination with respect to the observer, $\sim$
70\deg.

A qualitative explanation of the model comes from the schematic
precessing disc model shown in Fig.~3 by \citet{herx1:gerend76apj}.
In all cases, the main contribution to the optical flux is due to the
X-ray reprocessed emission from the disc, represented in the model
that fits the OM light curves by the average magnitude $M$. The angle
from which the observer views the pattern of re-emitted light from the
disc gradually changes through the precession cycle. A second
contribution to the optical flux arises from the heated face of the
companion. A maximum contribution of optical light is expected when
the observer sees the part of the star facing the disc, $\phi$ = 0.5,
and a minimum when the heated face is out of view, at
$\phi$ = 0.0. This component is represented in our model by the sine
component. A secondary minimum in the optical light curve is due to
the shadowing by the disc of the heated face at $\phi$ = 0.5. Depending
on the orientation of the disc with respect to the star, such minima
could be prominent, as in Obs~3, or almost invisible, as in
Obs~5. This secondary minimum is also visible in the average light
curve in \citet{1254:motch87apj}.

In \src, dipping should preferentially occur when the inclination of
the disc on the line of sight is highest. The disc inclination is never
large enough to produce total eclipses, but at orbital phase $\sim$
0.8, the bulge, assumed to be fixed in phase on the disc edge, can partially
block the view to the neutron star at some precessional phases. In the
case of Her~X-1, dipping episodes would correspond to the short "on"
state occurring between phases 0.4 and 0.65 in the ephemeris of
\citet{herx1:gerend76apj}. For this particular disc orientation, the
1.7~d light curve of Her~X-1 displays a shoulder at orbital phase
$\sim$0.8 similar to that observed in \src. This additional light
component is probably due to the X-ray heated stellar region whose
centroid is facing the observer, an area unshielded by the disc in
this particular geometrical observer-disc configuration. Because of
the varying height of the bulge, the appearance of X-ray dips may not
necessarily follow a strictly period pattern. This possibly accounts
for the fast dip to no dip regime change sometimes observed
\citep[e.g.][]{1254:courvoisier86apj}. Non dipping episodes should coincide
with low disc inclinations which in the light curve of Her~X-1
correspond to precession phases $\sim$0 \citep{herx1:gerend76apj}. At
these precessional phase, the disc casts a maximum shadow on the
secondary star at inferior conjunction thus producing a secondary
minimum at orbital phase 0.5 in Her~X-1, similar to that seen in the
light curve of Obs~3. Finally, shallow dipping would occur at an intermediate
configuraiton between the deep dipping and the non dipping conditions, when
the the bulge is observed at grazing angles due to its varying height.

Although disc precession can qualitatively explain the appearance and
disappearance of X-ray dips as well as the gross features of the
optical light curve, we stress that such an explanation remains
unproven in the absence of detailed light curve modelling. In
particular, the compatibility of the disc opening angle, bulge height
and possible tilt angles with the observed light curve cannot be
ascertained.

Finally, an additional effect may be needed to explain the enhanced dipping
activity during two years of the ASM light curves.

\begin{acknowledgements}
Based on observations obtained with XMM-Newton, an ESA science
mission with instruments and contributions directly funded by ESA
member states and the USA (NASA) and INTEGRAL, an ESA project
with instruments and science data centre funded by ESA member
states (especially the PI countries: Denmark, France, Germany,
Italy, Switzerland, Spain), Czech Republic and Poland and with the
participation of Russia and the USA.
This research has made use of 
data obtained from the High Energy Astrophysics Science Archive 
Research Center (HEASARC), provided by NASA's Goddard 
Space Flight Center.
\end{acknowledgements}

%------------------------------------
%       References
%------------------------------------

\bibliographystyle{aa}
\bibliography{1254}

\end{document}